\documentclass[conference]{IEEEtran}
\input{setup.tex}
\newacronym{stmv}{STMV}{Satellite Tobacco Mosaic Virus}
\newacronym{lqcd}{LQCD}{Lattice Quantum-Chromodynamics}
\newacronym{hpl}{HPL}{High-Performance Linpack}
\newacronym{fft}{FFT}{Fast Fourier Transform}

\makeglossaries

\title{Application-Driven Exascale: The JUPITER Benchmark Suite}

\author{
  \IEEEauthorblockN{%
  Andreas Herten\,\orcidlink{0000-0002-7150-2505}, %
  Sebastian Achilles\,\orcidlink{0000-0002-1943-6803}, %
  Damian Alvarez\,\orcidlink{0000-0002-8455-3598}, %
  Jayesh Badwaik\,\orcidlink{0000-0002-5252-8179}, %
  Eric Behle\,\orcidlink{0000-0001-6616-8867}, %
  Mathis Bode\,\orcidlink{0000-0001-9922-9742},\\ %
  Thomas Breuer\,\orcidlink{0000-0003-3979-4795}, %
  Daniel Caviedes-Voulli\`eme\,\orcidlink{0000-0001-7871-7544}, %
  Mehdi Cherti\,\orcidlink{0000-0001-5865-0469}, %
  Adel Dabah\,\orcidlink{0000-0001-9175-469X}, %
  Salem El Sayed\,\orcidlink{0000-0002-7217-6027}, \\
  Wolfgang Frings\,\orcidlink{0000-0002-9463-9264}, %
  Ana Gonzalez-Nicolas\,\orcidlink{0000-0003-2869-8255}, %
  Eric B.~Gregory\,\orcidlink{0000-0001-9408-5719}, %
  Kaveh Haghighi Mood\,\orcidlink{0000-0002-8578-4961}, %
  Thorsten Hater\,\orcidlink{0000-0002-6249-7169},\\ %
  Jenia Jitsev\,\orcidlink{0000-0002-1221-7851}, %
  Chelsea Maria John\,\orcidlink{0000-0003-3777-7393}, %
  Jan H.~Meinke\,\orcidlink{0000-0003-2831-9761}, %
  Catrin I.~Meyer\,\orcidlink{0000-0002-9271-6174}, %
  Pavel Mezentsev\,\orcidlink{0009-0000-4114-5482}, %
  Jan-Oliver Mirus\,\orcidlink{0009-0006-7975-1393}, \\
  Stepan Nassyr\,\orcidlink{0000-0002-0035-244X}, %
  Carolin Penke\,\orcidlink{0000-0002-4043-3885}, %
  Manoel R\"{o}mmer\,\orcidlink{0009-0007-3513-5932}, %
  Ujjwal Sinha\,\orcidlink{0000-0002-4609-0940}, %
  Benedikt von St. Vieth\,\orcidlink{0000-0002-5386-633X}, %
  Olaf Stein\,\orcidlink{0000-0002-6684-7103}, \\
  Estela Suarez\,\orcidlink{0000-0003-0748-7264}, %
  Dennis Willsch\,\orcidlink{0000-0003-3855-5100}, %
  Ilya Zhukov\,\orcidlink{0000-0003-4650-3773} %
  }
  \IEEEauthorblockA{
    \textit{J\"{u}lich Supercomputing Centre} \\
    \textit{Forschungszentrum J\"{u}lich} \\
    J\"{u}lich, Germany\\
    }
}

\begin{document}

\maketitle
\thispagestyle{fancy}
\lhead{}
\rhead{}
\chead{}
\lfoot{\footnotesize{SC24, November 17-22, 2024, Atlanta, Georgia, USA\newline 979-8-3503-5291-7/24/\$31.00 \copyright 2024 IEEE}}
\rfoot{}
\cfoot{}
\renewcommand{\headrulewidth}{0pt}
\renewcommand{\footrulewidth}{0pt}
\pagestyle{plain}  %

\begin{abstract}

Benchmarks are essential in the design of modern HPC installations, as they define key aspects of system components. Beyond synthetic workloads, it is crucial to include real applications that represent user requirements into benchmark suites, to guarantee high usability and widespread adoption of a new system. Given the significant investments in leadership-class supercomputers of the exascale era, this is even more important and necessitates alignment with a vision of Open Science and reproducibility.
In this work, we present the JUPITER Benchmark Suite, which incorporates 16 applications from various domains. It was designed for and used in the procurement of JUPITER, the first European exascale supercomputer. We identify requirements and challenges and outline the project and software infrastructure setup. We provide descriptions and scalability studies of selected applications and a set of key takeaways. The JUPITER Benchmark Suite is released as open source software with this work at \href{https://github.com/FZJ-JSC/jubench}{\url{github.com/FZJ-JSC/jubench}}. 

\end{abstract}
\begin{IEEEkeywords}
Benchmark, Procurement, Exascale, System Design, System Architecture, GPU, Accelerator
\end{IEEEkeywords}

\section{Introduction}
The field of High Performance Computing (HPC) is governed by the interplay of capability and demand driving each other forward. 
During the design and purchase phase of supercomputer procurements for public research, the capability of a machine is usually assessed not only by theoretical, system-inherent numbers, but also by effective numbers relating to actual workloads. These workloads are traditionally benchmark programs that test specific aspects of the system design --- like the floating-point throughput, memory bandwidth, or internode latency. While these \emph{synthetic benchmarks} are well-suited for the assessment of distinct features, for a more integrated and realistic perspective, they should be complemented by \emph{application benchmarks}. Application benchmarks use state-of-the-art scientific applications to assess the performance of integrated designs. Complex application profiles utilize various types of hardware resources dynamically during the benchmark's runtime, showcasing real-world strengths and limitations of the system.

This paper introduces the \emph{JUPITER Benchmark Suite}, a comprehensive collection of 23 benchmark programs meticulously documented and designed to support the procurement of JUPITER, Europe’s first exascale supercomputer. On top of 7 synthetic benchmarks, 16 application benchmarks were developed in close collaboration with domain scientists to ensure relevance and rigor. Additionally, this paper offers valuable insights into the state-of-the-practice of exascale procurement, shedding light on the challenges and methodologies involved.

Preparations for the procurement of JUPITER were launched in early 2022 and finally came to fruition with the awarding of the contract in October 2023. JUPITER is funded by the EuroHPC Joint Undertaking (\qty{50}{\percent}), Germany's Federal Ministry for Education and Research (\qty{25}{\percent}), and the Ministry of Culture and Science of the State of North Rhine-Westphalia of Germany (\qty{25}{\percent}), and is hosted at Jülich Supercomputing Centre (JSC) of Forschungszentrum Jülich. As part of the procurement, the benchmark suite was developed to motivate the system design and evaluate the proposals committed for the Request for Proposals. The suite focuses on application benchmarks to ensure high practical usability of the system. This work presents the \emph{JUPITER Benchmark Suite} in detail, highlighting design choices and project setup, describing the benchmark workloads, and releasing them as open source software.
The suite includes 23 benchmarks across different domains, each with unique characteristics such as compute-intensive, memory-intensive, and I/O-intensive workloads. The applications are grouped into three categories: Base, representing a mixed base workload for the system, High-Scaling, highlighting scalability to the full exascale system, and synthetic, determining various key hardware design features. 
The benchmark suite represents a first step towards \emph{Continuous Benchmarking} to detect system anomalies during the production phase of JUPITER. 

The main contributions of this paper are:
\begin{itemize}
    \item An in-depth description of the use of benchmarks in HPC procurement, including relevant background information. 
    \item The description of a novel methodology to assess exascale system designs in the form of \emph{High-Scaling benchmarks}.
    \item The development of suitable benchmark workloads based on a representative set of scientific problems, applications, and synthetic codes. 
    \item Scalability results on the preparation system for all application benchmarks of the suite. 
    \item Insights and best practices learned from application scaling and the procurement process in general.
    \item Release of the full JUPITER Benchmark Suite as open source software.
\end{itemize}

After providing background information (\autoref{sec:background}) and details on the used infrastructure (\autoref{sec:infrastructure}), the suite's individual benchmarks are presented in \autoref{sec:applications}. Key takeaways are provided in \autoref{sec:lessons}, followed by a conclusion and outlook in \autoref{sec:conclusion}.

\section{Background}
\label{sec:background}

\subsection{Requirements}
\label{sec:requirements}
The main requirements of the benchmark suite stem from the need to represent existing and upcoming user communities in the system design process. This ensures a fitting design and fosters adoption of the system by users. The suite must cover the wide user portfolio of the HPC center, containing typical applications from various domains utilizing the current HPC infrastructure, and also represent expected future workloads. Moreover, it is essential to ensure diversity in terms of methods, programming languages, and execution models, since such diversity is an inherent characteristic of the application portfolio for upcoming large-scale systems. 

The context of a procurement poses high requirements for \emph{replicability}, \emph{reproducibility}, and \emph{reusability}~\cite{Fehr16RRR}. Replicability, i.e., the seamless execution on the same hardware by the developer, is an elementary requirement to guarantee robustness. Beyond that, reproducibility describes the seamless execution on different hardware by someone else, making it a key requirement in the context of the procurement since both the site and the system provider must be able to run the suite to obtain the same results. Ensuring reusability, i.e., designing the framework for easy adaptation for a variety of tasks in the future, is essential to justify the substantial investment involved in the creation of the suite. 

Vendors participating in the procurement process must also invest considerable time in system-specific adjustments while meeting the procurement's requirements and complying to its rules, usually within short time scales. Therefore, it is in everyone's best interest to have clear, well-defined guidelines and, ideally, to leverage an existing benchmark suite to build on established expertise.

\subsection{JUPITER Procurement Scheme}
The procurement for the JUPITER system uses a Total-Cost-of-Ownership-based (TCO) value-for-money approach, in which the number of executed reference workloads over the lifespan of the system determines the value. Given the size of state-of-the-art supercomputers as well as their corresponding power consumption, costs for electricity and cooling are a substantial part of the overall project budget. Using a mixture of application benchmarks as well as synthetic benchmarks, operational costs are computed in a well-defined procedure. While synthetic benchmarks allow for assessing key performance characteristics, they do not allow a realistic assessment of resource consumption during the lifetime of the system for TCO. Therefore, a greater emphasis is placed on application performance rather than on synthetic tests. 

Given the targeted system performance of \qty[per-mode=symbol]{1}{\exa\floppersec} with \qty{64}{\bit} precision, an additional novel benchmark type focusing on the scale of the system is introduced --- \emph{High-Scaling} benchmarks. A subset of applications able to fully utilize JUPITER is identified and use cases were defined to make them part of this novel category. In the course of the paper, we will refer to either \emph{High-Scaling} benchmarks or \emph{Base} benchmarks, to differentiate between both.

The JUPITER system is envisioned to consist of two connected sub-systems, \emph{modules} in the Modular Supercomputing Architecture (MSA) concept~\cite{etp4hpcMSA}. JUPITER Booster is the exascale-level module utilizing massive parallelism through GPUs for maximum performance with high energy efficiency (\unit{\flopperjoule})%
. JUPITER Cluster is a smaller general-purpose module employing state-of-the-art CPUs for applications with lower inherent parallelism and stronger memory demands. Both compute modules are procured jointly, together with a third module made of high-bandwidth flash storage. The benchmark suite has dedicated benchmarks for all modules, partly even benchmarks spanning Cluster and Booster, dubbed \emph{MSA} benchmarks. Execution targets of the benchmarks are listed in the last columns of \autoref{tab:allthebenchmarks}.

During the procurement, the results of the execution of the benchmark suite for a given system proposal are weighted and combined to compute a value-for-money metric. The outcomes are compared and incorporated with other aspects into the final assessment of the system proposals.

\subsection{Implementation for JUPITER}
\label{sec:design}

\begin{figure*}[t]
    \centering
    \includegraphics[width=0.7\textwidth]{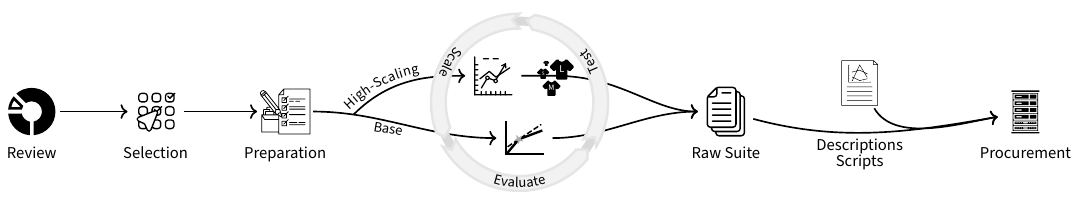} %
    \caption{Major steps in the creation of the JUPITER Benchmark Suite. Based on current workloads, a set of applications is selected. Benchmarks are prepared and then optimized in a feedback loop. Finally, descriptions are revised, and the suite is packaged for procurement.}
    \label{fig:jubench-flow}
    \vspace*{-3pt}
\end{figure*}

The previous two sections outline general requirements for the JUPITER Benchmark Suite. Their assessment and implementation is laid out in the following and are visually sketched in \figurename~\ref{fig:jubench-flow}.

Based on an analysis of current and previous compute-time allocations on predecessor systems, the suite covers a variety of scientific areas and includes applications from the domains of Weather and Climate, Neuroscience, Quantum Physics, Material Design, Biology, and others. Beyond that, diversity in workloads is realized: Artificial-Intelligence-based (AI) methods as well as classical simulations, codes based on C/C++ and Fortran, OpenACC and CUDA. Various application profiles are included, such as memory-bound or sparse computations. Future trends of workloads, e.g., the uptake of machine learning algorithms, are inferred from general trends in research communities and from recent changes in allocations on the predecessor system. 

The requirement of replicability is met by streamlining and automating benchmark execution employing the JUBE framework~\cite{jube}. Reproducibility is ensured by extensive documentation of all components and the verification of computational results. Thorough testing ensures stable execution in different environments, e.g., with a varying number of compute devices, and lowers risks of commitments by vendors. Reusability is accomplished by a modular design, effective project management workflows, clear licensing, and open source publication. To ensure a high quality of the suite, the benchmarks are standardized through a well-defined setup, with consistent directory structures, uniform descriptions, and similar JUBE configurations. The infrastructure aspects are discussed further in \autoref{sec:infrastructure}).

For each of the Base benchmarks, i.e., the sixteen benchmarks used for the TCO/value-for-money calculation, a Figure-of-Merit (FOM) is identified and normalized to a time-metric. In most cases, the FOM is the runtime of either the full application or a part of it. In case the application focuses on rates, the time-metric is achieved by pre-defining the number of iterations and multiplying with the rate. 
Each benchmark is executed on a certain number of nodes of the preparation system (see \autoref{sec:jwb}) in order to create a reference execution time. The number of nodes is usually selected to be 8, but deviations are possible due to workload-inherent aspects. This is a trade-off between resource economy and agility in benchmark creation (fewer nodes are more productive), and robustness towards anticipated generational leaps in the hardware of the envisioned system (fewer nodes incorporate latent danger to severely change application profiles when the requested runtime can be achieved with one node%
). 
The time-metric, determined on the reference number of nodes, is the value to be improved upon and committed to by proposals of system designs. The number of nodes used to surpass the time-metric can be freely specified by the proposal, but is typically smaller than the reference number of nodes.

Five applications used in the Base benchmarks are capable of scaling to the full scale of the preparation system. They form the additional \textit{High-Scaling} category and define a set of benchmarks that aim to compare the proposed system designs with the preparation system for large-scale executions. By requirement, the future system is known to achieve $\qty{1}{\exa\floppersec}$ \gls{hpl} performance, implying a theoretical peak performance larger than that.  
In JUPITER’s procurement, runtimes must be committed for the High-Scaling benchmarks using a $\qty[qualifier-mode = bracket]{1}{\exa\floppersec\theoretical}$ ($\qty[qualifier-mode = bracket]{1}{\exa\floppersec}$ theoretical peak performance) sub-partition. For each high-scaling application, a workload is defined to fill a \qty[qualifier-mode = bracket]{50}{\peta\floppersec\theoretical} sub-partition of the preparation system (about $640$ nodes) and a $20\times$ larger sub-partition of the future system ($20 \times \qty{50}{\peta\floppersec} = \qty{1}{\exa\floppersec}$).\footnote{Some benchmarks have algorithmic limitations, like requirement of powers-of-two in node counts. In this case, the smaller, closest compatible number of nodes is taken (for example, $512$ nodes).}  The final assessment is based on the ratio of the runtime value committed for the future \qty[qualifier-mode = bracket]{1}{\exa\floppersec\theoretical} sub-partition and the reference value. 
When the benchmark’s specific workload configuration fills up a large portion of the GPU memory on the preparation system, there is a danger that on a proposed system the  scaled-up version could become limited by available memory and not showcase an accelerator’s full compute capability. This is due to the trend of growing imbalance between the advancement of compute power and memory. To give more flexibility in system design, up to four reference variants of the respective workload are prepared, taking up \qty{25}{\percent} (tiny, T), \qty{50}{\percent} (small, S), \qty{75}{\percent} (medium, M),  and \qty{100}{\percent} (large, L) of the available GPU memory on the preparation system (\qty{40}{\giga\byte}), respectively. The system proposal may choose the variant that best exploits the available memory on the proposed accelerator after scale-up.

The seven synthetic benchmarks are selected to test individual features of the hardware components, such as compute performance, memory bandwidth, I/O throughput, and network design. Each benchmark has an individual FOM with unique rules and possibly sub-benchmarks, evaluated distinctly.

\subsection{Related Work}

Benchmarks have always played an important role in HPC. One objective is the assessment and comparison of technical solutions, whether at the scale of world-leading cluster systems~\cite{top500,Dongarra03HPL} or on a smaller node-level scale \cite{Sorokin20,Xu15NAS, Bienia08PARSEC}. Benchmarks can be specific to one application topic~\cite{Herten19ICEI, Albers22beNNch, Farrell21MLPerf} or cover multiple areas in the form of suites \cite{Dongarra12HPCChallenge,Stratton12Parboil,Danalis10SHOC,Che09Rodinia}.

Foundational work focused on identifying common patterns across various applications, with the objective of classifying them based on similar computational and communication characteristics \cite{asanovic06dwarfs}. These patterns, referred to as \emph{dwarfs}, are defined at a high level of abstraction and are intended to capture the most relevant workloads in high-performance computing. \autoref{tab:dwarfs} categorizes the benchmarks of the JUPITER Benchmark Suite according to these dwarfs, and also gives the predominant scientific domain a benchmark represents.  While dwarfs can be viewed as a set of blueprints for application-inspired synthetic benchmarks, their simplicity by design prevents them to fully capture all dynamics of real applications. To address this issue in supercomputer co-design, more complex and versatile computational motifs, termed \emph{octopodes}, are proposed by \citeauthor{MatsuokaRethinkingProxy}~\cite{MatsuokaRethinkingProxy}.

\begin{table}
    \centering
    \caption{%
        Relation of benchmarks of the JUPITER Benchmark Suite to domains\textsuperscript{§} and \emph{Berkeley Dwarfs}~\cite{asanovic06dwarfs}; other use-cases of the applications might have other profiles.
    }
    \label{tab:dwarfs}
\definecolor{oneColor}{HTML}{3f7389}
\definecolor{twoColor}{HTML}{66d6c9}
\definecolor{threeColor}{HTML}{DABFFF}
\definecolor{fourColor}{HTML}{E76F51}
\definecolor{fiveColor}{HTML}{E9C46A}
\definecolor{sixColor}{HTML}{B0D7FF}
\definecolor{sevenColor}{HTML}{E574BC}
\renewcommand{\checkmark}{$\cdot$}
\renewcommand{\checkmark}{$\bullet$}
\newcommand{\mymark}{\checkmark}
\newcommand{\markOne}{\textcolor{oneColor}{\mymark}}
\newcommand{\markTwo}{\textcolor{twoColor}{\mymark}}
\newcommand{\markThree}{\textcolor{threeColor}{\mymark}}
\newcommand{\markFour}{\textcolor{fourColor}{\mymark}}
\newcommand{\markFive}{\textcolor{fiveColor}{\mymark}}
\newcommand{\markSix}{\textcolor{sixColor}{\mymark}}
\newcommand{\markSeven}{\textcolor{sevenColor}{\mymark}}
\begin{threeparttable}[t]
\begin{tblr}{%
    colspec={ccccccc},
    font=\footnotesize, 
    row{1-Z}={font=\footnotesize},
    row{2}={font=\footnotesize\bfseries}, 
    vline{3}={2-Z}{dashed},
    rowsep=0pt,
    width=\linewidth,
    column{3-Z}={leftsep=5pt,rightsep=5pt},
    column{3}={leftsep=5pt},
    row{3}={abovesep=2pt},
}
    & & \SetCell[c=1]{l}{\myrot{\markOne\,Dense LA}} & \SetCell[c=1]{l}{\myrot{\markTwo\,Sparse LA}} & \SetCell[c=1]{l}{\myrot{\markThree\,Spectral}} & \SetCell[c=1]{l}{\myrot{\markFour\,Particle}} & \SetCell[c=1]{l}{\myrot{\markFive\,Structured Grid}} & \SetCell[c=1]{l}{\myrot{\markSix\,Unstructured Grid}} & \SetCell[c=1]{l}{\myrot{\markSeven\,Monte Carlo}} \\
    Benchmark   & Domain       & \SetCell[c=7]{c}{Dwarfs} \\
    \midrule
    Amber*      & MD           & & & \markThree & \markFour & & & \\ %
    Arbor       & Neurosci.    & \markOne & \markTwo & & & & \markSix & \\ %
    Chroma-QCD  & QCD          & & \markTwo& & &  \markFive & & \markSeven \\ %
    GROMACS     & MD           & & & \markThree & \markFour & & & \\ %
    ICON        & Climate      & & \markTwo & & & \markFive & & \\ %
    JUQCS       & QC           & & \markTwo & & & & & \\ %
    nekRS       & CFD          & & \markTwo & \markThree & & & \markSix & \\ %
    ParFlow*    & Earth Sys.   & & \markTwo & & & \markFive & & \\ %
    PIConGPU    & Plasma       & & & & \markFour & & \markSix & \\ %
    Quantum Espresso & Materials Sci. & \markOne & & \markThree & \markFour & & & \\ %
    SOMA*       & Polymer Sys. & & & & \markFour & & & \markSeven  \\ %
    MMoCLIP     & AI (MM)      & \markOne & & & & & & \\ %
    Megatron-LM & AI (LLM)     & \markOne & & & & & & \\ %
    ResNet*     & AI (Vision)  & \markOne & & & & & & \\ %
    DynQCD      & QCD          & & \markTwo & & & \markFive & &  \markSeven\\ %
    NAStJA      & Biology      & & & & & \markFive & & \markSeven \\[1ex] %
    Graph500    & Graph        & \SetCell[c=7]{c}{\emph{Graph Traversal (D. 9)}}& & & & & & \\
    HPCG        & CG           & & \markTwo & & & & & \\ %
    HPL         & LA           & \markOne & & & & & & \\
    IOR         & Filesys.     & \SetCell[c=7]{c}{\emph{Input/Output}}& & & & & & &  \\ %
    LinkTest    & Network      & \SetCell[c=7]{c}{\emph{P2P, Topology}}& & & & & & & \\ %
    OSU         & Network      & \SetCell[c=7]{c}{\emph{Message Exchange, DMA}}& & & & & & & \\ %
    STREAM      & Memory       & \SetCell[c=7]{c}{\emph{Regular Access}}& & & & & &\\
    \end{tblr}
    \begin{tablenotes}
        \item[*] The benchmarks were prepared for the procurement, but not actually used.
        \item[§] The following abbreviations are used: 
        \emph{MD} - Molecular Dynamics; 
        \emph{QCD} - Quantum Chromo Dynamics; 
        \emph{QC} - Quantum Computing;
        \emph{CFD} - Computational Fluid Dynamics;
        \emph{MM} - Multi-Modal;
        \emph{LLM} - Large Language Model;
        \emph{LA} - Linear Algebra;
        \emph{P2P} - Point-to-Point;
        \emph{DMA} - Direct Memory Access.%
    \end{tablenotes}
    \end{threeparttable}
\end{table}

SPEC~\cite{Juckeland15SPEC, Boehm18SPEC} is one of the most extensive, well-known benchmarking initiatives aimed at commercial users. The benchmark suite \emph{SPECaccel2023} uses the offloading APIs OpenACC and OpenMP to measure performance with computationally intensive parallel applications, following the principle “same source code for all”. 
Benchmarks for the use case of system design and exascale procurement~\cite{Malaya23Frontier} have specific requirements (see \autoref{sec:requirements}) and are typically not made publicly available due to concerns regarding sensitive information and elaborate implementation. The PRACE Unified European Applications Benchmark Suite~\cite{ueabs2024} represents a step towards a culture of open sharing, but its future support is uncertain. Other notable efforts include the CORAL-2 benchmarks~\cite{coral2} used for procurement of the three exascale systems in the US, and the recently developed NERSC-10 Benchmark Suite~\cite{nersc10} used in an ongoing procurement. HPC centers can benefit from each other's experience, driven by a spirit of Open Science, reproducibility, and sustainable software development~\cite{Fehr16RRR}. The integration of DevOps principles, such as Continuous Benchmarking, is gaining popularity to support these aims~\cite{Pearce23Benchpark,Adamson24Jacamar}. 

\section{Benchmark Infrastructure}
\label{sec:infrastructure}

\colorlet{cluster}{fzjblue!70!white}
\colorlet{booster}{OliveDrab}  %
\colorlet{msa}{cluster!50!booster!50!fzjred}
\DeclareRobustCommand{\execBoosterGpu}{\tikz[baseline=(b.base)]{\node (b) [fill=booster, text=white, font=\sffamily] {B};\node at (b.south east) [circle, fill=booster, text=white, thick, font=\sffamily\tiny, inner sep=0.5pt] {G};}}
\DeclareRobustCommand{\execBoosterCpu}{\tikz[baseline=(b.base)]{\node (b) [fill=booster, text=white, font=\sffamily] {B};\node at (b.south east) [fill=booster, text=white, font=\sffamily\tiny, inner sep=1.2pt] {C};}}
\DeclareRobustCommand{\execClusterCpu}{\tikz[baseline=(b.base)]{\node (b) [fill=cluster, text=white, font=\sffamily] {C};\node at (b.south east) [fill=cluster, text=white, font=\sffamily\tiny, inner sep=1.2pt] {C};}}
\DeclareRobustCommand{\execClusterBooster}{\tikz[baseline=(b.base)]{\node (b) [fill=msa, text=white, font=\sffamily] {M};\node (gc) at (b.south east) [text=white, font=\sffamily\tiny, inner sep=1.2pt] {GC};\fill [fill=msa, rotate=90] let \p1 = ($(gc.south west)-(gc.north west)$) in ([yshift=-2pt]gc.north west) arc (0:180:{veclen(\x1, \y1)/2}) -- (gc.south east) -- (gc.north east) -- cycle;\node (gc) at (b.south east) [text=white, font=\sffamily\tiny, inner sep=1pt] {GC};}}

\newcommand{\checkBoosterGpu}{\textcolor{booster}{\checkmark}}
\newcommand{\checkBoosterCpu}{\textcolor{booster}{\checkmark}}
\newcommand{\checkClusterCpu}{\textcolor{cluster}{\checkmark}}
\newcommand{\checkClusterBooster}{\textcolor{msa}{\checkmark}}
\newcommand{\inlinify}[1]{\resizebox{!}{1.6ex}{#1}}

\newcommand{\tableheader}{&\bfseries\makecell{Benchmark\\Name} & \bfseries\makecell{Progr. Language,\\{}[Libraries, ]Prog. Models} & \bfseries Licence & \bfseries\makecell{Nodes\\Base} & \bfseries\makecell{Nodes\\High-Scale} & \multicolumn{4}{c}{\bfseries \makecell{Module/\\Device}}}
\newcommand{\tablesubheader}{& & & & & $N^\text{Mem Vars}$ & \execBoosterGpu & \execBoosterCpu & \execClusterCpu & \execClusterBooster}
\begin{table*}
    \centering
\begin{threeparttable}[t]
    \caption{Overview of components of the JUPITER Benchmark Suite. Some defining details are given. If used, significant libraries are shown (all benchmarks use MPI for distribution). For \textbf{High-Scale} benchmarks, the super-script indicates the available memory variants (tiny, small, medium, large). For \textbf{Module/Device}, the following abbreviations are used: \inlinify{\execBoosterGpu} -- Execution on GPUs of JUPITER Booster, \inlinify{\execBoosterCpu} -- Execution on the CPUs of JUPITER Booster, \inlinify{\execClusterCpu} -- Execution on the CPUs of JUPITER Cluster, \inlinify{\execClusterBooster} -- MSA execution with CPUs of JUPITER Cluster and GPUs of JUPITER Booster.}
    \label{tab:allthebenchmarks}
    \centering
    \begin{tabular}{c@{\hspace{8pt}}cccclc@{\hspace{1pt}}c@{\hspace{1pt}}c@{\hspace{1pt}}c}
        \toprule
        \multicolumn{4}{c}{\bfseries Application Features} & \multicolumn{6}{c}{\bfseries Execution Targets}\\
        \cmidrule(r){1-4} \cmidrule(r){5-10}
        \tableheader \\
        \tablesubheader \\
        \midrule
        \tikzmark{app start}
            & Amber* & Fortran, CUDA & \emph{Custom} & 1 & & \checkBoosterGpu \\
        & Arbor & C++, CUDA/HIP & BSD-3-Clause & 8 & $642^\text{T,S,M,L}$ & \checkBoosterGpu\\
        & Chroma-QCD & C++, QUDA, CUDA/HIP & JLab & 8 & $512^\text{S,M,L}$ & \checkBoosterGpu \\
        & GROMACS & C++, CUDA/SYCL & LGPLv2.1 & 3/128 & & \checkBoosterGpu \\
        & ICON & Fortran/C, OpenACC/CUDA/HIP & BSD-3-Clause & 120/300 & & \checkBoosterGpu \\
        & JUQCS & Fortran, CUDA/OpenMP & \emph{None} & 8 & $512^\text{S,L}$ & \checkBoosterGpu & & & \checkClusterBooster\\
        & nekRS & C++/C, OCCA, CUDA/HIP/SYCL & BSD-3-Clause & 8 & $642^\text{S,M,L}$ & \checkBoosterGpu\\
        & ParFlow* & C, Hypre, CUDA/HIP & LGPL & 4 & & \checkBoosterGpu \\
        & PIConGPU & C++, Alpaka, CUDA/HIP & GPLv3+ & 8 & $640^{\text{S,M,L}}$ & \checkBoosterGpu\\
        & Quantum Espresso & Fortran, ELPA, OpenACC/CUF & GPL & 8 & & \checkBoosterGpu\\
        & SOMA* & C, OpenACC & LGPL & 8 & & \checkBoosterGpu \\
        & MMoCLIP & Python, PyTorch, CUDA/ROCm\tnote{1} & MIT & 8 & & \checkBoosterGpu \\
        & Megatron-LM & Python, PyTorch/Apex, CUDA/ROCm\tnote{1} & BSD-3-Clause & 96 & & \checkBoosterGpu \\
        & ResNet* & Python, TensorFlow, CUDA/ROCm\tnote{1} & Apache-2.0 & 10 & &\checkBoosterGpu\\
        & DynQCD & C, OpenMP & \emph{None} & 8 & & & & \checkClusterCpu \\
        \tikzmark{app end} 
            & NAStJA & C++, MPI & MPL-2.0 & 8 & & & & \checkClusterCpu \\
        \tikzmark{syn start}
            & Graph500 & C, MPI & MIT & 4/16/all & &  & \checkBoosterCpu \\ 
        & HPCG & C++, OpenMP, CUDA/HIP & BSD-3-Clause & 1/4/all &  & \checkBoosterGpu & & \checkClusterCpu \\ 
        & HPL & C, BLAS, OpenMP, CUDA/HIP & BSD-4-Clause & 1/16/all &  & \checkBoosterGpu & & \checkClusterCpu \\ 
        & IOR & C, MPI & GPLv2 & -/$>64$\tnote{§} &  & & \checkBoosterCpu & \checkClusterCpu \\ 
        & LinkTest & C++, MPI/SIONlib & BSD-4-Clause+ & all &  & & \checkBoosterCpu & \checkClusterCpu & \checkClusterBooster \\ 
        & OSU & C, MPI, CUDA & BSD & 1/2 &  & \checkBoosterGpu & \checkBoosterCpu & \checkClusterCpu \\ 
        \tikzmark{syn end}
            & STREAM & C, CUDA/ROCm/OpenACC & \emph{Custom} & 1 &  & \checkBoosterGpu & & \checkClusterCpu \\
        \midrule
        \tablesubheader \\
        \tableheader\\
        \bottomrule
    \end{tabular}
    \begin{tikzpicture}[overlay, remember picture]
        \path [|-|, draw=gray, shorten >=-0.3ex, shorten <=-1.3ex] (pic cs:app start) to node [fill=white, midway, rotate=90, text=gray, font=\itshape] {Application} (pic cs:app end);
        \path [|-|, draw=gray, shorten >=-0.3ex, shorten <=-1.3ex] (pic cs:syn start) to node [fill=white, midway, rotate=90, text=gray, font=\itshape] {Synthetic} (pic cs:syn end);
    \end{tikzpicture}
    \begin{tablenotes}
        \item[*] The benchmarks were prepared for the procurement, but not actually used.
        \item[1] For PyTorch and TensorFlow, CUDA and ROCm backends are available; through \emph{extensions}, also backends for Intel GPUs exist (not in mainline repositories).
        \item[§] IOR features two sub-benchmarks, \emph{easy} and \emph{hard}. The number of nodes is a free parameter in easy. In hard, it can also be chosen freely, as long as more than 64 nodes are taken.
    \end{tablenotes}
\end{threeparttable}
\vspace*{-3pt}
\end{table*}

\subsection{Preparation Systems}
\label{sec:jwb}
The JUPITER Benchmark Suite was prepared on JUWELS, in particular JUWELS Booster, a top 20 system~\cite{top500nov2023} hosted at JSC~\cite{juwels}. JUWELS Booster was installed in 2020 and provides a performance of \qty[qualifier-mode = bracket]{73}{\peta\floppersec\theoretical} peak and \qty{44}{\peta\floppersec} for the \gls{hpl}. The system is connected to the JUST~5 storage system~\cite{graf2021just}. 
JUWELS Booster provides 936 GPU nodes integrated into 39 Eviden BullSequana XH2000 racks, with 2~racks (48 nodes) building a \emph{cell} in the DragonFly+ topology of the high-speed interconnect. Each node has 4~NVIDIA A100 GPUs and 4~NVIDIA Mellanox InfiniBand HDR200 adapters, with one adapter available per GPU. 2 AMD EPYC Rome 7402 CPUs ($2 \times 24$ cores) are connected to \qty{512}{\giga\byte} DDR4 memory.

Preparations for the High-Scaling benchmarks utilized a \qty[qualifier-mode = bracket]{50}{\peta\floppersec\theoretical} sub-partition of the JUWELS Booster. %

JUWELS provides general software dependencies through EasyBuild~\cite{hoste2012easybuild}. Reproducibility is achieved by either using upstream installation recipes, \emph{easyconfigs}, or upstreaming custom recipes.%

\subsection{JUBE}
\label{sec:jube}
Every benchmark is implemented in the JUBE~\cite{jube} workflow environment to facilitate productive development and reproducibility. In benchmark-specific definition files, \emph{JUBE scripts}, parameters and execution steps (compilation, computation, data processing, verification) are defined. These are then interpreted by the JUBE runtime, resolving dependencies and eventually submitting jobs for execution to the batch system. By inheriting from system-specific definition files, \emph{\texttt{platform.xml}}, batch submission templates are populated and independence of the underlying system is achieved. The various sub-benchmarks and variants are implemented by \emph{tags}, which select different versions of parameter definitions. After execution, the benchmark results are presented by JUBE in a concise tabular form, including the FOM.

Within the JUPITER procurement, the JUBE scripts are part of the documentation. They exactly define execution parameters and instructions with descriptive annotations. A textual documentation is provided as part of the benchmark-accompanying description.

\subsection{Descriptions}
Beyond the execution reference through JUBE scripts, each benchmark is accompanied by an extensive description. All descriptions are normalized, using identical structure with similar language. Example parts are information about the source and the compilation, execution parameters and rules, detailed instructions for execution and verification, sample results, and concluding commitment requests. In all relevant sections, relations to the JUBE scripts are made in addition to the textual descriptions. For the vendors, the use of JUBE is recommended but not mandatory.

PDFs generated from the benchmark descriptions are part of the committed procurement documentation, including hashes of archived benchmark repositories.

\subsection{Git and Submodules}
All components of a benchmark are available in a single Git repository as a single source of truth. A common structure is established, containing description, JUBE scripts, auxiliary scripts, benchmark results, and the source code of the benchmarked application. Utilizing the attached issue tracker, project management and collaboration are facilitated.

Per default, the sources are included as references in the form of Git Submodules. Submodules enable a direct linkage to well-defined versions of source code, but do not unnecessarily clutter the benchmark repository by static, potentially extensive copies. They are well integrated into the Git workflow and easily updated. In cases where inclusion as a Git Submodule is not possible, scripts and detailed instructions for download are provided. %

For delivery as part of the procurement specification package, each benchmark repository is archived as a tar file.

If too large for inclusion in the Git repository, input data is provided as a separate download, including a verifying hash.

\subsection{Project Management}

The benchmark suite development efforts were supported efficiently by clear project management workflows over several months. 

A core team of  HPC specialists and scientific researchers initiated the process early, curating a list of potential benchmarks based on their expertise and experience with existing HPC systems, while also incorporating insights from previous procurements. In close collaboration with domain scientists, this list was gradually refined to ensure a balanced and diverse selection that met the requirements outlined in \autoref{sec:requirements}.

Competent teams, each led by a team captain, were responsible for individual applications. GitLab issues were used to document biweekly meetings and track per-application progress in the form of a pre-defined checklist with 11 points (ranging from source code availability, over JUBE integration, to description creation). Collaborative hack days facilitated collaboration while running the benchmarks on the preparation system.

\section{Benchmarks}
\label{sec:applications}

This section describes the JUPITER Benchmark Suite, containing 23 benchmarks: 7 synthetic and 16 application benchmarks. In the procurement process, the number of application benchmarks was reduced to 12. 
\autoref{tab:allthebenchmarks} gives an overview of all benchmarks, including application features and execution targets. %

With this work, the JUPITER Benchmark Suite is released as open source software at \url{https://github.com/FZJ-JSC/jubench}, with individual repositories for each benchmark~\cite{zenodo_amber,zenodo_arbor,zenodo_chroma,zenodo_gromacs,zenodo_icon,zenodo_juqcs,zenodo_nekrs,zenodo_parflow,zenodo_picongpu,zenodo_qe,zenodo_soma,zenodo_mmoclip,zenodo_megatron,zenodo_resnet,zenodo_dynqcd,zenodo_nastja,zenodo_graph500,zenodo_hpcg,zenodo_hpl,zenodo_icon,zenodo_linktest,zenodo_osu,zenodo_stream,zenodo_streamgpu}.

\subsection{Application Benchmarks}
In the following, we describe in detail eleven of the application benchmarks, divided into Base and High-Scaling benchmarks. The benchmarks are based on prominent workloads in the HPC community and were specifically developed for the suite. Given their complex computational dynamics, a certain level of technical and scientific background is necessary and will be provided accordingly.  The remaining benchmarks, Amber, ParFlow, SOMA, ResNet, and DynQCD, are briefly introduced first for completeness; they are either using closed-source software (DynQCD) or were ultimately not used for the JUPITER procurement (the others).

\begin{itemize}
    \item \emph{Amber}~\cite{AmberGPU,AmberTools} is a popular commercial molecular dynamics code for biomolecules. The \gls{stmv} case from the Amber20 benchmark suite~\cite{AmberBenchmark} (\num{1067095} atoms) is chosen. The code is mainly optimized for single GPU calculations and is not intended to scale beyond a single node.
    \item \emph{ParFlow}~\cite{ashby1996parallel,jones2001newton} is a massively-parallel, open source, integrated hydrology model for surface and subsurface flow simulation. The ClayL test from ParFlow's test suite (simulating infiltration into clay soil) is selected, with a problem size of $1008\times1008\times240$ cells~\cite{Hokkanen2021}.
    \item \emph{SOMA}~\cite{soma} performs Monte Carlo simulations for the “Single Chain in Mean Field” model~\cite{daoulas2006single}, studying the behaviour of soft coarse-grained polymer chains in a solution.  
    \item \emph{ResNet}~\cite{he2015deep} uses convolutions and residual connections for training deep neural networks and serves as a reference model in computer vision tasks. The suite includes ResNet50, implemented in TensorFlow with Horovod.
    \item \emph{DynQCD}~\cite{dynqcd} is a CPU-only code which performs numerical simulations for \gls{lqcd}. %
    The benchmark generates 600 quark propagators using a conjugate gradient solver for sparse \gls{lqcd} fermion matrices, with high demands to the memory sub-system.
\end{itemize}

\subsubsection{Base Benchmarks (Selection)}

\begin{figure*}[t]
    \centering
    \includegraphics[width=\textwidth]{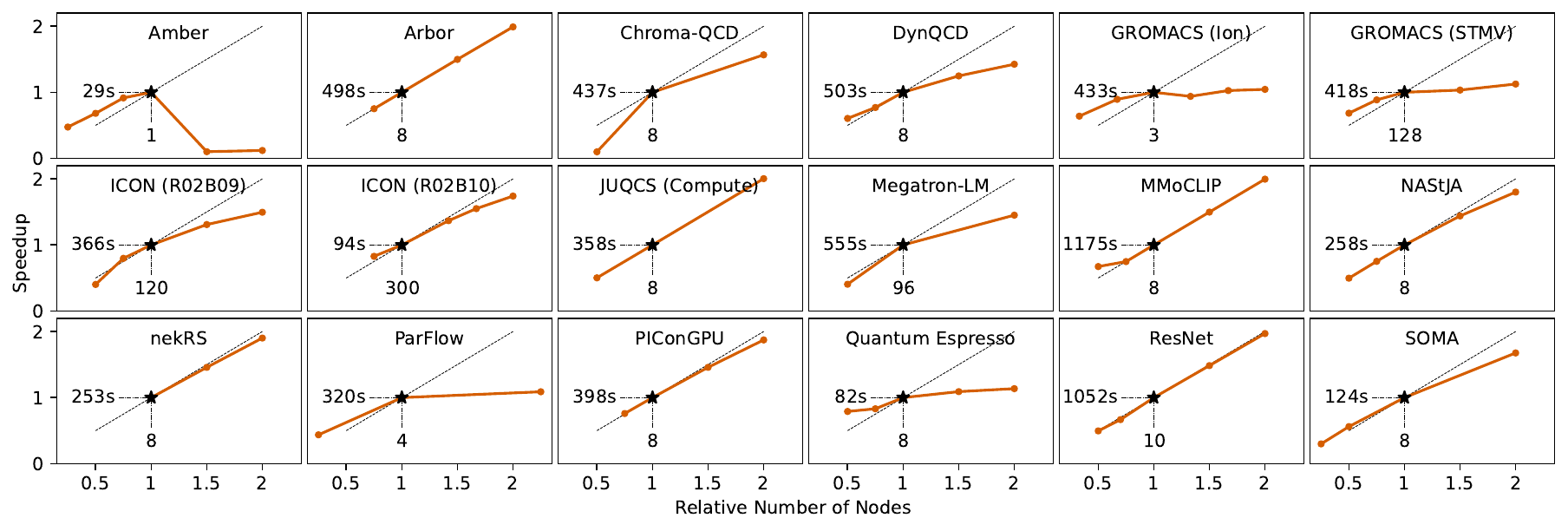}
    \caption{Overview of relative runtimes of all Base applications on the reference system, JUWELS Booster. Shown at $(1,1)$ is the execution on the reference number of nodes with the reference runtime, with the absolute values shown close to the horizontal and vertical axes, respectively. Beyond the reference execution, strong-scaled relative runtimes (with respect to the reference runtime) on the surrounding number of nodes are given (usually $0.5\times$, $0.75\times$, $1.5\times$, and $2\times$ the reference; some benchmarks deviate). As an example, consider Arbor: the reference number of nodes ($8$) is noted at horizontal $1$, the reference runtime of \qty{498}{\second} at vertical $1$; further data-points are given for $4$ nodes (\qty{663}{\second}), $12$ nodes (\qty{332}{\second}), and $16$ nodes (\qty{250}{\second}) -- $0.5\times$, $1.5\times$, and $2\times$ the reference number of nodes of $8$. See also \autoref{tab:allthebenchmarks}.}
    \label{fig:tco}
    \vspace*{-3pt}
\end{figure*}

The Base benchmarks are designed to incentivize system designs that optimize the time to solution. They are first executed on a reference number of nodes on the preparation system (see \autoref{sec:design}). Figure~\ref{fig:tco} gives an overview of application runtimes and respective strong scaling behaviors for surrounding number of nodes. While the absolute number can be used to judge system designs quantitatively, the strong scaling behavior can be used as an additional data point to understand the overall design qualitatively. Note the example for reading the graph in the figure caption.

\paragraph{GROMACS}
GROMACS~\cite{Berendsen1995,pall_tackling_2014,Abraham2015,Pall2020} is a versatile package to perform molecular dynamics simulations, focusing on biochemical molecules and soft condensed matter systems. The application integrates Newton's equations of motion for systems with hundreds to millions of particles and provides time-resolved trajectories. 
Two biological systems from the Unified European Applications Benchmark Suite ({UEABS})~\cite{ueabs2024} are used, test cases \emph{A} and \emph{C}. Test case A simulates a GluCl ion channel embedded in a membrane. Test case C contains 27 replicas of the \gls{stmv} with about \num{28000000} atoms and allows testing the scalability of system-supplied \gls{fft} libraries.

\paragraph{ICON}
The ICOsahedral Non-hydrostatic model (ICON)~\cite{Zaengl2015} is a modelling framework for weather, climate, and environmental prediction used for operational weather forecasting at the German Weather Service. ICON also provides an Earth System Model for climate simulations, i.e., a general circulation model of the atmosphere, including a land module~\cite{Giorgetta2018} and an ocean model~\cite{Korn2022}. While the atmosphere part has been ported to GPUs~\cite{Giorgetta2022}, the ocean component is still running on CPUs only. %
ICON is available under a permissive open source licence.
The JUPITER benchmark case is based on the atmosphere component, %
with global forecast simulation in two resolutions, resulting in two sub-benchmarks: R02B09 (\qty{5}{\kilo\meter} grid point distance) and R02B10 (\qty{2.5}{\kilo\meter} grid point distance)~%
\cite{Hohenegger2023}. The coarser resolution is targeted for execution on 120 nodes, and the finer resolution is for 300 nodes. While reasonable scaling to $2\times$ the node count (240 nodes and 600 nodes, respectively) is possible, this is not the usual mode of operation for ICON.
These simulations are crucial for ICON's development towards a storm-resolving climate model with \qty{1}{\kilo\meter} resolution or even less~\cite{Stevens2020}. A unique aspect of the ICON benchmark is its large input dataset: R02B09 requires 
\qty{1.8}{\tera\byte} of data, R02B10 needs \qty{4.5}{\tera\byte}. Therefore, the ICON benchmark also tests the performance of I/O operations on a system.

\paragraph{Megatron-LM} %
Megatron-LM~\cite{shoeybi2020megatronlm} is a prominent codebase in Natural Language Processing (NLP), known for its vast scale and performance capabilities. It employs the transformer model architecture~\cite{vaswani2023attention} and leverages various parallelization techniques and optimizations
\cite{narayanan2021efficient,korthikanti2022reducing,dao2023flashattention2,rajbhandari2020zero} through PyTorch to achieve high hardware utilization with excellent efficiency. The training of various recent open source GPT-like models was carried out with Megatron-LM~\cite{John2023OGX}, making this benchmark crucial to assess a system's capability to handle disruptive generative AI workloads.
The benchmark trains a 175 billion parameter model, converting the usual throughput metric (tokens per time) to a hypothetical time-to-solution FOM by training 20 million tokens.

\paragraph{MMoCLIP}

Contrastive Language-Image Pre-training (CLIP)~\cite{radford2021learning} is a method that conducts self-supervised learning on weakly-aligned image-text pairs with open vocabulary language, resulting in language-vision foundation models. The approach enables usage of substantially increased datasets, like web-scale datasets~\cite{Schuhmann2021, Schuhmann2022}. The OpenCLIP~\cite{ilharco_gabriel_2021_5143773, Cherti2023} codebase, an open source implementation of CLIP, enables efficiently distributed training of CLIP models by using multiple data parallelism schemes through PyTorch, scaling to more than a thousand GPUs. 
Due to its strong transferability and robustness, OpenCLIP is used in diverse multi-modal learning approaches and downstream applications~\cite{ldm,liu2023visual,sun2023emu}, and efficient training is crucial for the machine learning community dealing with open, fully reproducible foundation models.

The MMoCLIP benchmark is curated from OpenCLIP. It trains an ViT-L-14 model on a synthetic dataset of \num{3200000} image-text pairs and records the total training time as a FOM.

\paragraph{Quantum ESPRESSO}
Quantum ESPRESSO (QE)~\cite{QE1,QE2,QEGPU} is an open source, density-functional-theory-based electronic structure software used both in academia and industry. %
QE calculates different material properties using a plane wave basis set and pseudo-potentials and can exploit novel accelerators well~\cite{QEexa}.
The dominant kernel in QE performs a three-dimensional \gls{fft}, which is usually a memory-bound kernel and is communication-bound for large systems~\cite{QEexa}.

For the benchmark suite, the \emph{Car-Parrinello Molecular Dynamics} model was chosen. The benchmark is based on a use case created in the MaX project~\cite{MaXrepo,MaX} and does calculations for a slab of ZrO2 with 792 atoms.

\paragraph{NAStJA}

The Neoteric Autonomous Stencil code for Jolly Algorithms (NAStJA) is a massively-parallel simulation framework of biological tissues using a Cellular Potts Model~\cite{BerghoffNastja,Graner1992}. This model relies on nearest neighbour interactions and is parallelized by dividing the overall workload into multiple sub-regions, called blocks. Each block is treated independently by an MPI process, with boundaries being exchanged. Using NAStJA, tissues composed of thousands to millions of cells can be simulated at subcellular resolution~\cite{berghoff2020cells}. 
As a test case, adhesion-driven cell sorting is used, a common process in tissue development and segregation~\cite{Steinberg1962}.
The benchmark investigates the first 5050 Monte Carlo (MC) steps of a system of size $720\times720\times1152\,\unit{\micro\meter\cubed}$, containing roughly \num{600000} cells. %
NAStJA utilizes MPI for parallelization/distribution and is one of the few CPU-only benchmarks in the suite. The application exhibits an irregular memory access pattern at each iteration, which is not suitable for GPU execution. %

\subsubsection{High-Scaling Benchmarks}

In this section, we describe the five High-Scaling benchmarks in detail.

\begin{figure}[t]
    \centering
    \includegraphics[width=\columnwidth]{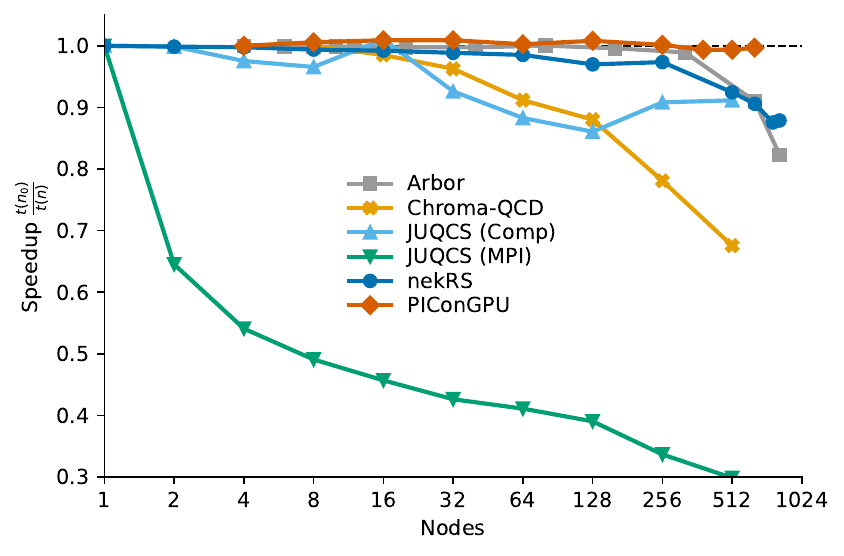}
    \caption{Weak scaling efficiency of the five High-Scaling benchmarks over a wide range of JUWELS Booster node numbers. For JUQCS, two lines are drawn; one for the computation and one for the communication (see section~\ref{app:juqcs}).}
    \label{fig:highscale}
\end{figure}
Figure~\ref{fig:highscale} juxtaposes the weak-scaling behaviours of the applications over a wide range of node numbers, using the reference HPC system JUWELS Booster.

\paragraph{Arbor}
Arbor is a library for simulating biophysically-realistic neural networks, bridging the gap between point and nanoscale models~\cite{arbor}. 
Developed in the HBP~\cite{hbp}, it aims at efficient use of modern HPC hardware behind an intuitive interface.
Neurons are modelled by morphology, ion channels, and connections. 
Arbor is written in C++ with a Python interface and available under a permissive open source license.
The user-centric description is discretized and aggregated to optimize data layouts for individual hardware. 
At runtime, the \emph{cable equation} is integrated alternating with a system of ODEs for the channels. 
Users model channels via a domain-specific language that must be compiled for the target hardware.
Communication is performed, concurrently with time evolution, every $n$ steps, determined by neural delays.
The benchmark is parameterized to fill the GPU memory in the variants T, S, M, L.
To differentiate from point models, it is weighted heavily towards computation, emphasized by a sparse connectivity.
A complex cell from the Allen Institute was selected and adapted to random morphologies of fixed depth~\cite{allen}. 
Cells are organized into rings propagating a single spike. 
Rings are interconnected to place load on the network without altering dynamics, yielding a deterministic, scalable workload.  
Profiling shows two cost centers: \qty{52}{\percent} ion channels and \qty{33}{\percent} cable equation; hiding communication completely. 
The Base version runs on 32 JUWELS Booster nodes, filling all 4 GPUs' memory.
This was scaled up to the full Booster to verify efficient resource usage and extrapolated to \qty{1}{\exa\flop}.
The number of generated spikes is used for validation.

\paragraph{Chroma}
Chroma~\cite{Edwards:2004sx} is an all-purpose application for \gls{lqcd} computations. It is compiled on top of the USQCD software stack~\cite{usqcd}, which provides \gls{lqcd}-specific libraries for communication, data-parallelism, I/O, and, importantly, sparse linear solver libraries optimized for different architectures. 
Key libraries used in this benchmark include QMP for MPI wrapping, QDP-JIT for data-parallelism and parallel-I/O via QIO, and the GPU-targeted QUDA solver library~\cite{Clark:2009wm}.
Chroma and the USQCD stack are open source and community-developed. Chroma is one of the most widely used LQCD suites and is representative of \gls{lqcd} codes in general.

\gls{lqcd} calculations generally depend heavily on solving very large, regular, sparse linear systems (dimension $10^6-10^9$ generally). Due to the regularity of the data and the calculation, \gls{lqcd} lends itself to many levels of concurrency.%

The Chroma \gls{lqcd} benchmark in the JUPITER Benchmark Suite 
contains the representative benchmarks for the Hybrid Monte-Carlo (HMC) component of the \gls{lqcd} simulations. In the benchmark, a number of HMC update trajectories are performed using the $3+1$ flavours of Clover Wilson fermions --- three light quark flavours with identical mass, and a fourth flavour with heavier mass --- and the L\"{u}scher-Weisz gauge action. The 4D lattice is initialized with a random $SU(3)$ element on each link. Checkpointing is disabled by a source patch to remove the I/O overhead for the calculations.
The benchmark also contains a fix to Chroma, allowing simulation of 4D lattice volumes greater than $2^{31}$ and among the largest \gls{lqcd} simulations anywhere to date. The benchmark performance is sensitive to the decomposition configuration used for distributing the 4D lattice to different tasks and to the affinity between the CPU, GPU, NUMA domains, and the network controller for each task.

The benchmark is validated by comparing the output with a reference solution with a tolerance of $10^{-10}$ for the Base benchmark and $10^{-8}$ for High-Scaling benchmarks.

The relevant metric (FOM) is the total time spent in HMC updates, excluding the first update, which includes overhead for tuning QUDA parameters. So a minimum of two updates must be prescribed.

\paragraph{JUQCS}
\label{app:juqcs}

JUQCS is a massively parallel simulator for universal gate-based Quantum Computers (QCs)~\cite{DeRaedt2007JUQCS} written in Fortran~90 using MPI, OpenMP, and CUDA. %
During the past decades, JUQCS has been used to benchmark some of the largest supercomputers worldwide, including the Sunway TaihuLight and the K computer~\cite{DeRaedt2019JUQCS} as well as JUWELS Booster~\cite{Willsch2022JUQCSG}, and it was part of Google's quantum supremacy demonstration~\cite{Google2019QuantumSupremacy}. Various versions of JUQCS are available in binary form as part of a container~\cite{JUQCSDocker}; a light version with available sources was created specifically for this benchmark suite~\cite{JUQCSLight}. 
JUQCS simulates an $n$-qubit gate-based QC by iteratively updating a rank-$n$ tensor of $2^n$ complex numbers (state vector) stored in double precision and distributed over the supercomputer's memory. %
The total available memory determines the size of the largest QC that can be simulated. For instance, a universal simulation of $n=45$ qubits requires a little over $16\times 2^{45}\,\unit{\byte} = \qty{0.5}{\pebi\byte}$. Many operations require the transfer of half of all memory, i.e.,~$2^n/2$ complex double-precision numbers, across the network, which can help to assess the performance of a supercomputer's communication network~\cite{DeRaedt2019JUQCS,Willsch2022JUQCSG}. 
As \figurename~\ref{fig:highscale} shows, the deviation of JUQCS w.r.t.~the theoretically expected linear scaling (green triangles) %
reveals both a drop in performance from intra-node to inter-node GPU communication (from 1 to 2 nodes) and another drop when communication enters the large-scale regime at 256 nodes. %

All present JUQCS benchmarks simulate successive applications of a single-qubit quantum gate that requires large memory transfers. The Base benchmark simulates $n=36$ qubits requiring $\qty{1}{\tebi\byte}$ of GPU memory. The High-Scaling benchmark contains two memory variants: a large memory variant with $n=42$ qubits requiring $\qty{64}{\tebi\byte}$ (\emph{L}) and a small memory variant with $n=41$ qubits requiring $\qty{32}{\tebi\byte}$ of GPU memory (\emph{S}). Rules are given for extrapolation to an Exascale setup using $n=46$ (L) or $n=45$ qubits (S). For all test cases, verification is done using the theoretically known results~\cite{DeRaedt2019JUQCS}. %

In addition, an MSA version of the JUQCS benchmark simulates $n=34$ qubits on both  JUWELS Cluster and Booster simultaneously. %
The total amount of memory is split into two parts, with $\qty{128}{\gibi\byte}$ residing on the CPU nodes and $\qty{128}{\gibi\byte}$ residing on the GPU nodes. MPI is used for communication between the Cluster and the Booster, and the number of MPI tasks is similarly split into two. On the Cluster, each MPI task launches 12 OpenMP threads, with one thread per CPU core. On the Booster, each MPI task controls one of the GPUs.

\paragraph{nekRS}

nekRS~\cite{Fischer2022} is a fast CFD solver designed for GPUs that solves the low-Mach Navier-Stokes equations (NSEs), potentially coupled with multiphysics effects. %
nekRS has been run at scale on many large supercomputers, featuring excellent 
time-to-solution due to its high GPU throughput, and was nominated for the 2023 Gordon-Bell Award~\cite{Merzari2023}. nekRS uses high-order spectral elements~\cite{Patera1984} in which the solution, data, and test functions are represented as locally structured $N^{\mathrm{th}}$-order tensor product polynomials on a set of $E$ globally unstructured curvilinear hexahedral brick elements. 
There are two key benefits to this strategy. First, high-order polynomial expansions significantly reduce the number of unknowns ($n\approx EN^3$) needed to achieve engineering tolerances. Second, the locally structured forms allow tensor product sum factorization, which yields low $\mathcal{O}(n)$ storage cost and $\mathcal{O}(nN)$ work complexity~\cite{Orszag1980}. 
The leading order $\mathcal{O}(nN)$ work terms can be cast as small dense matrix-matrix products with good computational intensity~\cite{Deville2002}. %
nekRS is written in C++ and the kernels are implemented using the portable Open Concurrent Compute Abstraction (OCCA) library \cite{occa} for abstraction between different parallel languages/hardware architectures.

The benchmark case is derived from a Rayleigh-Bénard convection (RBC) application~\cite{Samuel2024,Bode2024} which simulates turbulence induced by a temperature gradient --- a typical case executed at scale.%
The simulation domain is a \emph{sheet}. It is much more extended in the periodic directions than in the wall-bounded direction. The chosen polynomial order is $9$ with 600 time steps per run. Verification is based on pre-computed results and derived tolerances. 
The High-Scaling benchmark variants use between \num{28836900} (\textit{small}, \num{\sim 11229} per GPU) %
and \num{57760000} (\textit{large}, \num{\sim 22,492} per GPU) elements, which is more than the minimum number of elements required for the "strong scaling limit" of $\num{7000} - \num{8000}$ elements per GPU. The Base benchmark case uses \num{719104} elements resulting in \num{22472} elements per GPU. %

\paragraph{PIConGPU}

PIConGPU is an open source, fully relativistic particle-in-cell (PIC) code designed for studying laser-plasma interactions and astrophysical phenomena. It uses the PIC algorithm with several key components, namely, particle initialization, charge calculations using grid interpolation, field calculations using densities, and time-marching  due to Lorentz force. This approach allows particles to interact via fields on the grid rather than direct pairwise interactions, reducing computational steps from $N^2$ to $N$ for $N$ particles. 
PIConGPU employs a unique data model with asynchronous data transfers to handle the computational challenge. It can simulate complex plasma systems with billions of particles on GPU clusters~\cite{PIConGPU2013}. PIConGPU  is developed with a hardware-agnostic approach using the Alpaka library~\cite{ZenkerAsHES2016,Worpitz2015}, providing outstanding performance across all supported platforms, like CPUs, AMD and NVIDIA GPUs, and FPGAs~\cite{PIConGPU_alpaka}. 
The benchmark suite uses a 3D test-case simulating the Kelvin-Helmholtz Instability (KHI), a non-relativistic shear-flow instability, utilizing a pre-ionized hydrogen plasma with periodic boundary conditions. 
While relevant for various research communities, the nature of shear-flows and the use of periodic boundary conditions does not impose a significant load imbalance throughout the simulation. Therefore, the performance of the code is based on its structure rather than the physics of the problem. %
In the KHI use case, the number of particles per cell is kept constant to 25, using as many cells as the GPU memory allows. A grid size of $\vec{x} = (4096, 2048, 1024)$ is chosen for the small memory variant, and extended to $(4096, 2048, 2048)$ (M) and $(4096, 4096, 2560)$ (L) for the larger variants. PIConGPU employs domain decomposition for distribution, dividing the computational domain into smaller subdomains along three dimensions. To distribute along these three dimensions, the maximum number of nodes that can be utilized is limited to $640$, rather than $642$.

\subsection{Synthetic Benchmarks}
\label{sec:syn}
The JUPITER Benchmark Suite also includes seven well-known synthetic benchmarks: Graph500, HPCG, HPL, IOR, LinkTest, OSU, and STREAM (including a GPU variant). 
The IOR and LinkTest implementation are presented in the following, highlighting some unique aspects of the setup.

\paragraph{IOR}
IOR is the de facto standard for measuring I/O performance and is being used by the IO500~\cite{kunkel:2018:io500} to compare the I/O characteristics of storage systems. The benchmark provides a large list of parameters such as block size, transfer size, API, and task reordering, which in turn allows simulating multiple I/O patterns. 
To target the high-bandwidth, NVMe-based JUPITER storage module, the upper and lower bounds on the mean read and write bandwidth are our focus in the Benchmark Suite. Similar to IO500, two variants of the IOR are implemented, \emph{Easy} and \emph{Hard}. The Easy variant requires a transfer size of \qty{16}{\mebi\byte}, with each process writing to its own file. The Hard variant uses a transfer size of \qty{4}{\kibi\byte} and a block size of \qty{4}{\kibi\byte}, with all processes writing and reading a single file. The setup forces multiple processes to write to the same file system data block, stressing the filesystem with the lock processes. The remaining parameters were selected to avoid caching effects.
The number of nodes is a free parameter (with a lower bound) to allow for optimization on the level of parallelism that the underlying file system can provide.

\paragraph{LinkTest}
LinkTest~\cite{LinkTest} is designed to test point-to-point connections between processes in serial or parallel mode and is capable of handling very large numbers of processes%
. It is an essential tool in network operations, used mostly internally by system administrators for acceptance testing, maintenance, and troubleshooting.

The JUPITER Benchmark Suite utilizes LinkTest's \emph{bisection} test, to concisely evaluate interconnectivity between parts of the system's network, quantified by a single metric. In the bisection test, a number of test processes (one per high-speed network adapter) is separated to two equal halves of the system, and messages are bounced between partnering processes in parallel (bidirectional mode).  To achieve optimal bandwidth, the message size is set to \qty{16}{\mebi\byte}. An assessment is made mainly based on the minimum bisection bandwidth. %

\section{Lessons Learned}
\label{sec:lessons}

We developed the JUPITER Benchmark Suite building upon our experience from previous HPC system procurements.
The suite constitutes a substantial expansion from those earlier endeavours, and should be considered as a living object that will continue to evolve over time.
In the following, we summarize the lessons learned, covering first the perspective of application developers, then the benchmark suite creation, and finally the overall procurement process.

\subsection{Application Development}

The applications of the JUPITER Benchmark Suite not only need to be executed on current large-scale resources like JUWELS Booster, but also need to be extrapolated to larger future resources, amplifying scaling challenges. %

To understand the performance characteristics on a future system better, it proved useful for some application developers to create \textbf{models} of their applications. JUQCS, for example, has a non-trivial weak scaling behaviour. In the benchmark, the execution time is reported in relation to an ideal scenario, enabling comparability. A model was developed for nekRS to predict the performance of a later part of the simulation early in the process, allowing much shorter and more resource-efficient benchmarks. During scalability studies for PIConGPU, a model for the scaling behaviour could be developed, informing valid simulation parameters for the benchmark setup.

While the approximate scale of the future system is known, the details of the setup are not.
When domain decomposition is important for performance, preliminary studies are usually employed to determine the best parameters for production runs. But \textbf{decomposition studies} are impossible in the benchmark context, especially for an unknown system design. Through labour- and resource-intensive investigation, estimates, rules, or scripts for ideal domain decomposition were devised, e.g., for Chroma-QCD, PIConGPU, NAStJA, and DynQCD. This also documented the experience of individual researchers, improving reproducibility.
To understand different scaling regimes of the application, a network communication model was developed for JUQCS. The model can be employed to understand topological aspects of the high-speed network of current and future systems, for example with respect to congestion.

The preparation for \textbf{future system designs} had a direct effect on application development. For example, it became apparent for Arbor developers that they need to optimize memory usage, as memory capacity and bandwidth will continue to extend more slowly compared to compute performance. During benchmark preparation, they also needed to trade highly-valued user experience for scalability, as the approach of referring to connection endpoints with labels did not scale as required. A short-term solution (using local indexing) was found for the suite, and a hash-based solution is being developed upstream. On a similar note, Chroma-QCD authors needed to patch the code to facilitate execution on the envisioned scale.
Unexpected effects can occur depending on the extrapolation method to the future system. For Chroma-QCD, it was found out that the employed benchmark is not guaranteed to converge, and a cut-off after a certain number of iterations is a more robust approach.

Result \textbf{verification} is essential for a benchmark suite to ensure the validity of submitted results. Yet, the experience with verification during the suite preparation varied.
Some results could be verified either exactly (JUQCS), or within a certain numerical limit by comparing to a pre-computed solution (Chroma-QCD); more involved simulations were verified by extracting key metrics from the computed solution for comparison to a model (ICON, nekRS). The verification of some applications with iterative algorithms, which were stopped before convergence, relied on framework-inherent verification and required key data in the output (PIConGPU, Megatron-LM) --- arguably the weakest form of verification.

\subsection{Benchmark Design}

Creating a vast benchmark suite that picks up the status quo in workloads, and bringing it to mature levels, is a \textbf{resource-intensive endeavour}. Beyond the human resources, compute resources of the reference system constitute a significant investment. To use these resources efficiently, it is important to design benchmarks with runtimes as short as possible, while keeping it as large as needed. Short runtimes also enable swift turn-around times for rapid prototyping -- especially useful in a large suite. The size of the input data and the files generated at runtime should be minimized to ensure that the benchmark suite is easy to handle. For reproducibility, non-core application parts like pre-processing or data-staging should be kept short. Modelling domain decomposition effects, also beyond typical production execution profiles, is further consuming resources.

Preparing for target system designs with \textbf{unknown details and scales} beyond available resources is a demanding task. Care should be taken to consider future hardware trends in the benchmark design. To explicitly accommodate different compute-to-memory ratios, up to four memory variants of benchmarks were introduced. On the preparation system, the memory variants can be used to study artificially-limited compute profiles and determine possible bottleneck shifts on future systems. With unknown hardware, algorithmic behaviours might shift as well, and iterations may not converge. A more robust approach is to not compute until convergence, but stop after a predetermined amount of iterations.

Some parameters in the benchmark suite are free to chose, like the number of nodes, others are fixed, for example simulation parameters. Thorough \textbf{execution rules} and modification guidelines determine the envisioned outcome and need to be developed as part of the suite. Beyond general rules, benchmarks may explicitly deviate to either loosen or tighten rules. Parameter validation should be part of the overall verification process; further extending on the importance of the verification task.

\subsection{Procurement}

Using application benchmarks in the procurement of HPC systems is essential to realistically represent user requirements when deciding the configuration of the future system.
An additional challenge was given by the particular system architecture of JUPITER, in which \textbf{two compute modules} of very disparate sizes, a small CPU-only module and an exascale GPU-accelerated module, are coupled together with a shared storage module. It took some discussions until finding the right number and balance between CPU and GPU benchmarks, which ended up being in the ratio of about 1:5. 

To formulate and develop the benchmarks, it proved fruitful to \textbf{collaborate closely} with the domain researchers intending to utilize the system. Established relationships in joint projects are especially productive, while it is more difficult for new user domains. 
A fundamental limitation of our approach is the \textbf{reliance on existing application codes} executed on current systems. By design, disruptive approaches are not well covered and a tendency to favour evolutionary technology is introduced. However, considering the effort associated with adopting new technologies among HPC users, this focus on incremental developments is justified. 
Still, predicting future system \textbf{usage trends} is crucial --- like the AI applications in the suite, which aim at representing a user domain expected to gain importance over time. However, the rapidly evolving software and algorithms in this domain make it hard to accurately estimate their future needs. It is therefore important to consider the most recent breakthroughs in AI beyond the HPC context, including also commercially-driven domains.

The \textbf{time window} for the development of the JUPITER Benchmark Suite was limited by the constraints of the procurement process. The endeavour started several months before the procurement, and required dedicated work by tens of people.  
Clear management structures and collaboration platforms were essential \textbf{tools for extensive collaboration}. In particular, transparent communication with all bidders was crucial, which was possible thanks to the dialogue phase that was part of the procurement. The suite development fostered collaboration, team-building, and knowledge-sharing. Code and environment optimizations were openly shared between benchmark developers and vendors, iteratively improving the benchmarks further. The suite itself is now open source and can \textbf{benefit the wider HPC community}.

\section{Conclusions and Future Work}
\label{sec:conclusion}

In this paper, we presented the JUPITER Benchmark Suite, which has been successfully employed in the procurement process of the exascale system JUPITER. The suite has served as a valuable tool in assessing HPC system performance during the procurement and beyond. The benchmark applications (see \autoref{sec:applications}) were chosen to represent the workloads on the future system after careful consideration of requirements and constraints. Bidders used this suite to test different technologies, put together their proposals, and prepare and commit the associated performance numbers. The procuring entity selected the vendor based on these values, together with multiple additional evaluation criteria. At the time of writing, the JUPITER system is being installed. The benchmark suite will be employed again during the acceptance procedure. %

The JUPITER Benchmark Suite lays a foundation to further extend and automate HPC system benchmarking. Facilitated by the reusable design of the suite, Continuous Benchmarking will be realized as future work, employing the CI/CD features of GitLab in conjunction with novel tools such as Jacamar~\cite{Adamson24Jacamar}.
Running the suite at regular intervals (e.g., after maintenances), we will ensure that the system does not see performance degradation over its lifetime or after updates. 
Application optimization for JUPITER will continue during the system deployment and installation phase, utilizing experiences gained and tools created. 
\doubleblind{The developed benchmarking framework will be employed and further enhanced within the JUPITER Research and Early Access Program (JUREAP).}
We will strive for further improvements regarding the reproducibility of individual benchmarks, including a focus on verification. Also, individual technical enhancements are in progress (for example, using git-annex for the large input data).

\section*{Acknowledgment}
\addcontentsline{toc}{section}{Acknowledgment}

Many benchmarks presented in this work have roots in projects that received funding from national and European sources; for example, Quantum Espresso (from the MaX project~\cite{maxcoe}) or nekRS (from the CoEC project~\cite{coec}). The authors thank the funding agencies for their support and the colleagues from the projects for being available for advice when designing the suite.

The authors would like to thank the following people for their contributions: Max Holicki and Yannik Müller, for their contributions to the LinkTest benchmark; Jonathan Windgassen and Christian Witzler for their support with the nekRS benchmark; Hans De Raedt for the discussions and help with the JUQCS benchmark.

The double-anonymous review of this publication was enabled by the \emph{Anonymous GitHub} service at \url{https://anonymous.4open.science/}. The authors thank the creator, also for the immediate, last-minute support.

Finally, the authors would like to thank EuroHPC JU, BMBF, MKW-NRW, and GCS for the supportive atmosphere, productive discussions, and for the financial support of the JUPITER project.

\printbibliography

@inproceedings{Schuhmann2022,
title={{LAION}-5{B}: An open large-scale dataset for training next generation image-text models},
author={Christoph Schuhmann and Romain Beaumont and Richard Vencu and Cade W Gordon and Ross Wightman and Mehdi Cherti and Theo Coombes and Aarush Katta and Clayton Mullis and Mitchell Wortsman and Patrick Schramowski and Srivatsa R Kundurthy and Katherine Crowson and Ludwig Schmidt and Robert Kaczmarczyk and Jenia Jitsev},
booktitle={Thirty-sixth Conference on Advances in Neural Information Processing Systems (NeurIPS), Datasets and Benchmarks Track},
year={2022},
volume={35},
pages={25278--25294},
url={https://openreview.net/forum?id=M3Y74vmsMcY}
}

@article{ashby1996parallel,
  title={A parallel multigrid preconditioned conjugate gradient algorithm for groundwater flow simulations},
  author={Ashby, Steven F and Falgout, Robert D},
  journal={Nuclear science and engineering},
  volume={124},
  number={1},
  pages={145--159},
  year={1996},
  publisher={Taylor \& Francis},
  doi={10.13182/NSE96-A24230}
}

@article{jones2001newton,
  title={{N}ewton--{K}rylov-multigrid solvers for large-scale, highly heterogeneous, variably saturated flow problems},
  author={Jones, Jim E and Woodward, Carol S},
  journal={Advances in water resources},
  volume={24},
  number={7},
  pages={763--774},
  year={2001},
  publisher={Elsevier},
  doi={10.1016/S0309-1708(00)00075-0}
}

@inproceedings{Cherti2023,
  title={Reproducible scaling laws for contrastive language-image learning},
  author={Cherti, Mehdi and Beaumont, Romain and Wightman, Ross and Wortsman, Mitchell and Ilharco, Gabriel and Gordon, Cade and Schuhmann, Christoph and Schmidt, Ludwig and Jitsev, Jenia},
  booktitle={Proceedings of the IEEE/CVF Conference on Computer Vision and Pattern Recognition},
  pages={2818--2829},
  year={2023}
}

@Article{Schuhmann2021,
  author  = {Schuhmann, Christoph and Vencu, Richard and Beaumont, Romain and Kaczmarczyk, Robert and Mullis, Clayton and Katta, Aarush and Coombes, Theo and Jitsev, Jenia and Komatsuzaki, Aran},
  journal = {Data-Centric AI Workshop NeurIPS, arXiv:2111.02114},
  title   = {Laion-400m: Open dataset of clip-filtered 400 million image-text pairs},
  year    = {2021},
}

@article{daoulas2006single,
  title={Single chain in mean field simulations: Quasi-instantaneous field approximation and quantitative comparison with {M}onte {C}arlo simulations},
  author={Daoulas, Kostas Ch and M{\"u}ller, Marcus},
  journal={The Journal of Chemical Physics},
  volume={125},
  number={18},
  year={2006},
  publisher={AIP Publishing},
  doi={10.1063/1.2364506}
}

@article{soma,
title = {Multi-architecture Monte-Carlo (MC) simulation of soft coarse-grained polymeric materials: SOft coarse grained Monte-Carlo Acceleration (SOMA)},
journal = {Computer Physics Communications},
volume = {235},
pages = {463-476},
year = {2019},
issn = {0010-4655},
doi = {https://doi.org/10.1016/j.cpc.2018.08.011},
url = {https://www.sciencedirect.com/science/article/pii/S0010465518303072},
author = {L. Schneider and M. M{\"u}ller},
}

@inproceedings{ldm,
  title={High-resolution image synthesis with latent diffusion models},
  author={Rombach, Robin and Blattmann, Andreas and Lorenz, Dominik and Esser, Patrick and Ommer, Bj{\"o}rn},
  booktitle={Proceedings of the IEEE/CVF Conference on Computer Vision and Pattern Recognition},
  pages={10684--10695},
  year={2022}
}

@inproceedings{
liu2023visual,
title={Visual Instruction Tuning},
author={Haotian Liu and Chunyuan Li and Qingyang Wu and Yong Jae Lee},
booktitle={Thirty-seventh Conference on Neural Information Processing Systems},
year={2023},
url={https://openreview.net/forum?id=w0H2xGHlkw}
}

@inproceedings{sun2023emu,
  title={Emu: Generative Pretraining in Multimodality},
  author={Sun, Quan and Yu, Qiying and Cui, Yufeng and Zhang, Fan and Zhang, Xiaosong and Wang, Yueze and Gao, Hongcheng and Liu, Jingjing and Huang, Tiejun and Wang, Xinlong},
  booktitle={The Twelfth International Conference on Learning Representations},
  year={2023}
}

@Article{DeRaedt2007JUQCS,
  author           = {Koen {De Raedt} and Kristel Michielsen and Hans {De Raedt} and Binh Trieu and Guido Arnold and Marcus Richter and {\relax Th}omas Lippert and Hiroshi Watanabe and Nobuyasu Ito},
  title            = {Massively parallel quantum computer simulator},
  doi              = {10.1016/j.cpc.2006.08.007},
  pages            = {121},
  volume           = {176},
  journal          = {Comput. Phys. Commun.},
  year             = {2007},
}

@Article{DeRaedt2019JUQCS,
  author    = {Hans {De Raedt} and Fengping Jin and Dennis Willsch and Madita Willsch and Naoki Yoshioka and Nobuyasu Ito and Shengjun Yuan and Kristel Michielsen},
  title     = {Massively parallel quantum computer simulator, eleven years later},
  journal   = {Comput. Phys. Commun.},
  year      = {2019},
  volume    = {237},
  pages     = {47 - 61},
  doi       = {10.1016/j.cpc.2018.11.005},
}

@Article{Google2019QuantumSupremacy,
  author           = {Arute, Frank and Arya, Kunal and Babbush, Ryan and Bacon, Dave and Bardin, Joseph C. and Barends, Rami and Biswas, Rupak and Boixo, Sergio and Brandao, Fernando G. S. L. and Buell, David A. and Burkett, Brian and Chen, Yu and Chen, Zijun and Chiaro, Ben and Collins, Roberto and Courtney, William and Dunsworth, Andrew and Farhi, Edward and Foxen, Brooks and Fowler, Austin and Gidney, Craig and Giustina, Marissa and Graff, Rob and Guerin, Keith and Habegger, Steve and Harrigan, Matthew P. and Hartmann, Michael J. and Ho, Alan and Hoffmann, Markus and Huang, Trent and Humble, Travis S. and Isakov, Sergei V. and Jeffrey, Evan and Jiang, Zhang and Kafri, Dvir and Kechedzhi, Kostyantyn and Kelly, Julian and Klimov, Paul V. and Knysh, Sergey and Korotkov, Alexander and Kostritsa, Fedor and Landhuis, David and Lindmark, Mike and Lucero, Erik and Lyakh, Dmitry and Mandr{\`a}, Salvatore and McClean, Jarrod R. and McEwen, Matthew and Megrant, Anthony and Mi, Xiao and Michielsen, Kristel and Mohseni, Masoud and Mutus, Josh and Naaman, Ofer and Neeley, Matthew and Neill, Charles and Niu, Murphy Yuezhen and Ostby, Eric and Petukhov, Andre and Platt, John C. and Quintana, Chris and Rieffel, Eleanor G. and Roushan, Pedram and Rubin, Nicholas C. and Sank, Daniel and Satzinger, Kevin J. and Smelyanskiy, Vadim and Sung, Kevin J. and Trevithick, Matthew D. and Vainsencher, Amit and Villalonga, Benjamin and White, Theodore and Yao, Z. Jamie and Yeh, Ping and Zalcman, Adam and Neven, Hartmut and Martinis, John M.},
  title            = {Quantum supremacy using a programmable superconducting processor},
  doi              = {10.1038/s41586-019-1666-5},
  pages            = {505--510},
  volume           = {574},
  journal          = {Nature},
  year             = {2019},
}

@Article{Willsch2022JUQCSG,
  author           = {Dennis Willsch and Madita Willsch and Fengping Jin and Kristel Michielsen and Hans {De Raedt}},
  title            = {{GPU}-accelerated simulations of quantum annealing and the quantum approximate optimization algorithm},
  doi              = {10.1016/j.cpc.2022.108411},
  pages            = {108411},
  volume           = {278},
  journal          = {Comput. Phys. Commun.},
  year             = {2022},
}

@Electronic{JUQCSDocker,
  author       = {{De Raedt}, Hans and Willsch, D.},
  title        = {J\"ulich Universal Quantum Computer Simulator (docker container)},
  url          = {https://jugit.fz-juelich.de/qip/juqcs-docker},
  howpublished = {\url{https://jugit.fz-juelich.de/qip/juqcs-docker.git}},
  year         = {2021},
}

@Electronic{JUQCSLight,
  author       = {{De Raedt}, Hans and Willsch, D.},
  title        = {J\"ulich Universal Quantum Computer Simulator (light version)},
  url          = {https://jugit.fz-juelich.de/qip/juqcs-light},
  howpublished = {\url{https://jugit.fz-juelich.de/qip/juqcs-light.git}},
  year         = {2024},
}

@Article{Fischer2022,
  author           = {P. Fischer and S. Kerkemeier and M. Min and Y. Lan and M. Phillips and T. Rathnayake and E. Merzari and A. Tomboulides and A. Karakus and N. Chalmers and T. Warburton},
  title            = {{NekRS, a GPU-accelerated spectral element Navier–Stokes solver}},
  pages            = {102982},
  volume           = {114},
  journal          = {Parallel Computing},
  year             = {2022},
  doi              = {10.1016/j.parco.2022.102982}
}

@inproceedings{Merzari2023,
author = {Merzari, Elia and Hamilton, Steven and Evans, Thomas and Min, Misun and Fischer, Paul and Kerkemeier, Stefan and Fang, Jun and Romano, Paul and Lan, Yu-Hsiang and Phillips, Malachi and Biondo, Elliott and Royston, Katherine and Warburton, Tim and Chalmers, Noel and Rathnayake, Thilina},
title = {Exascale Multiphysics Nuclear Reactor Simulations for Advanced Designs},
year = {2023},
isbn = {9798400701092},
publisher = {Association for Computing Machinery},
address = {New York, NY, USA},
doi = {10.1145/3581784.3627038},
booktitle = {Proceedings of the International Conference for High Performance Computing, Networking, Storage and Analysis},
articleno = {3},
numpages = {11},
location = {Denver, USA},
series = {SC '23}
}

@INPROCEEDINGS{arbor,
  author={Akar, Nora Abi and Cumming, Ben and Karakasis, Vasileios and Küsters, Anne and Klijn, Wouter and Peyser, Alexander and Yates, Stuart},
  booktitle={2019 27th Euromicro International Conference on Parallel, Distributed and Network-Based Processing (PDP)}, 
  title={Arbor — A Morphologically-Detailed Neural Network Simulation Library for Contemporary High-Performance Computing Architectures}, 
  year={2019},
  volume={},
  number={},
  pages={274-282},
  doi={10.1109/EMPDP.2019.8671560}}

@article{allen,
  title={Systematic integration of structural and functional data into multi-scale models of mouse primary visual cortex},
  author={Yazan N. Billeh and Binghuang Cai and Sergey L. Gratiy and Kael Dai and Ramakrishnan Iyer and Nathan W. Gouwens and Reza Abbasi-Asl and Xiaoxuan Jia and Joshua H. Siegle and Shawn R. Olsen and Christof Koch and Stefan Mihalas and Anton Arkhipov},
  publisher={Neuron},
  journal = {Neuron},
  volume = {106},
  number = {3},
  pages = {388-403.e18},
  year = {2020},
  issn = {0896-6273},
  doi = {https://doi.org/10.1016/j.neuron.2020.01.040}
}

@article{hbp,
  title={The {Human Brain Project}—Synergy between neuroscience, computing, informatics, and brain-inspired technologies},
  author={Amunts, Katrin and Knoll, Alois C and Lippert, Thomas and Pennartz, Cyriel MA and Ryvlin, Philippe and Destexhe, Alain and Jirsa, Viktor K and D’Angelo, Egidio and Bjaalie, Jan G},
  journal={PLoS Biology},
  volume={17},
  number={7},
  pages={e3000344},
  year={2019},
  publisher={Public Library of Science San Francisco, CA USA},
  doi={10.1371/journal.pbio.3000344}
}

@Article{Patera1984,
  author           = {A. T. Patera},
  title            = {A spectral element method for fluid dynamics: Laminar flow in a channel expansion},
  pages            = {468--488},
  volume           = {54},
  journal          = {Journal of Computational Physics},
  year             = {1984},
  doi              = {10.1016/0021-9991(84)90128-1}
}

@Book{Deville2002,
  author           = {M. O. Deville and P. F. Fischer and E. H. Mund},
  title            = {High-Order Methods for Incompressible Fluid Flow},
  publisher = {Cambridge Monographs on Applied and Computational Mathematics, Cambridge University Press},
  year             = {2002},
  doi              = {10.1017/CBO9780511546792}
}

@Article{Orszag1980,
  author           = {S. A. Orszag},
  title            = {Spectral methods for problems in complex geometries},
  pages            = {70--92},
  volume           = {37},
  journal          = {Journal of Computational Physics},
  year             = {1980},
  doi              = {10.1016/0021-9991(80)90005-4}
}

@Article{Samuel2024,
  %author           = {R. J. Samuel and M. Bode and J. D. Scheel and K. R. Sreenivasan and J. Schumacher},
  author           = {{Hidden (Blind Review)}},
  title            = {Boundary layers in thermal convection are fluctuation-dominated},
  journal          = {Journal of Fluid Mechanics (submitted)},
  year             = {2024},
}

@Article{Bode2024,
  %author           = {M. Bode and others},
  author           = {{Hidden (Blind Review)}},
  title            = {A High-Scaling In-Situ Workflow for Deciphering Boundary Layer Effects in High-{R}ayleigh-Number Convection},
  journal          = {(submitted)},
  year             = {2024},
}

@article{berghoff2020cells,
  title={Cells in Silico -- introducing a high-performance framework for large-scale tissue modeling},
  author={Berghoff, Marco and Rosenbauer, Jakob and Hoffmann, Felix and Schug, Alexander},
  journal={BMC bioinformatics},
  volume={21},
  number={1},
  pages={1--21},
  year={2020},
  publisher={Springer},
  doi={10.1186/s12859-020-03728-7}
}

@article{Graner1992,
  title = {Simulation of biological cell sorting using a two-dimensional extended {P}otts model},
  author = {Graner, Fran\c{c}ois and Glazier, James A.},
  journal = {Phys. Rev. Lett.},
  volume = {69},
  issue = {13},
  pages = {2013--2016},
  numpages = {0},
  year = {1992},
  month = {09},
  publisher = {American Physical Society},
  doi = {10.1103/PhysRevLett.69.2013}
}

@INPROCEEDINGS{BerghoffNastja,
  author={Berghoff, Marco and Kondov, Ivan},
  booktitle={2018 IEEE/ACM 9th Workshop on Latest Advances in Scalable Algorithms for Large-Scale Systems (scalA)}, 
  title={Non-collective Scalable Global Network Based on Local Communications}, 
  year={2018},
  volume={},
  number={},
  pages={25-32},
  doi={10.1109/ScalA.2018.00007}}

@INPROCEEDINGS{he2015deep,
  author={He, Kaiming and Zhang, Xiangyu and Ren, Shaoqing and Sun, Jian},
  booktitle={2016 IEEE Conference on Computer Vision and Pattern Recognition (CVPR)}, 
  title={Deep Residual Learning for Image Recognition}, 
  year={2016},
  volume={},
  number={},
  pages={770-778},
  doi={10.1109/CVPR.2016.90}}

@TECHREPORT{etp4hpcMSA,
      author       = {Suarez, Estela and Eicker, Norbert and Moschny, Thomas and
                      Pickartz, Simon and Clauss, Carsten and Plugaru, Valentin
                      and Herten, Andreas and Michielsen, Kristel and Lippert,
                      Thomas},
      title        = {{M}odular {S}upercomputing {A}rchitecture: {A} success
                      story of {E}uropean ${R}\&{D}$},
      number       = {9},
      publisher    = {ETP4HPC},
      reportid     = {FZJ-2022-03088, 9},
      series       = {White Paper},
      pages        = {24},
      year         = {2022},
      cin          = {JSC},
      cid          = {I:(DE-Juel1)JSC-20090406},
      pnm          = {5121 - Supercomputing $\&$ Big Data Facilities (POF4-512) /
                      5122 - Future Computing $\&$ Big Data Systems (POF4-512) /
                      DEEP-SEA - DEEP – SOFTWARE FOR EXASCALE ARCHITECTURES
                      (955606) / DEEP - Dynamical Exascale Entry Platform (287530)
                      / DEEP-ER - DEEP Extended Reach (610476) / DEEP-EST - DEEP -
                      Extreme Scale Technologies (754304)},
      pid          = {G:(DE-HGF)POF4-5121 / G:(DE-HGF)POF4-5122 /
                      G:(EU-Grant)955606 / G:(EU-Grant)287530 / G:(EU-Grant)610476
                      / G:(EU-Grant)754304},
      typ          = {PUB:(DE-HGF)3 / PUB:(DE-HGF)29},
      doi          = {10.5281/ZENODO.6508394}
}

@article{Steinberg1962,
author = {Malcolm S. Steinberg },
title = {On the mechanism of tissue reconstruction by dissociated cells, {I.} {P}opulation kinetics, differential adhesiveness, and the absence of directed migration},
journal = {Proceedings of the National Academy of Sciences},
volume = {48},
number = {9},
pages = {1577-1582},
year = {1962},
doi = {10.1073/pnas.48.9.1577}
}

@article{AmberGPU,
author = {Salomon-Ferrer, Romelia and Götz, Andreas W. and Poole, Duncan and Le Grand, Scott and Walker, Ross C.},
title = {Routine Microsecond Molecular Dynamics Simulations with {AMBER} on {GPU}s. 2. Explicit Solvent Particle Mesh {E}wald},
journal = {Journal of Chemical Theory and Computation},
volume = {9},
number = {9},
pages = {3878-3888},
year = {2013},
doi = {10.1021/ct400314y}
}

@article{AmberTools,
author = {Case, David A. and Aktulga, Hasan Metin and Belfon, Kellon and Cerutti, David S. and Cisneros, G. Andrés and Cruzeiro, Vinícius Wilian D. and Forouzesh, Negin and Giese, Timothy J. and Götz, Andreas W. and Gohlke, Holger and Izadi, Saeed and Kasavajhala, Koushik and Kaymak, Mehmet C. and King, Edward and Kurtzman, Tom and Lee, Tai-Sung and Li, Pengfei and Liu, Jian and Luchko, Tyler and Luo, Ray and Manathunga, Madushanka and Machado, Matias R. and Nguyen, Hai Minh and O’Hearn, Kurt A. and Onufriev, Alexey V. and Pan, Feng and Pantano, Sergio and Qi, Ruxi and Rahnamoun, Ali and Risheh, Ali and Schott-Verdugo, Stephan and Shajan, Akhil and Swails, Jason and Wang, Junmei and Wei, Haixin and Wu, Xiongwu and Wu, Yongxian and Zhang, Shi and Zhao, Shiji and Zhu, Qiang and Cheatham, Thomas E. III and Roe, Daniel R. and Roitberg, Adrian and Simmerling, Carlos and York, Darrin M. and Nagan, Maria C. and Merz, Kenneth M. Jr.},
title = {AmberTools},
journal = {Journal of Chemical Information and Modeling},
volume = {63},
number = {20},
pages = {6183-6191},
year = {2023},
doi = {10.1021/acs.jcim.3c01153}
}

@Electronic{AmberBenchmark,
  author       = {David Cerutti},
  title        = {Amber20 benchmarks},
  url          = {https://ambermd.org/Amber20_Benchmark_Suite.tar.gz},
  %howpublished = {\url{https://ambermd.org/Amber20_Benchmark_Suite.tar.gz},
  year         = {2020},
}

@article{QEexa,
author = {Carnimeo, Ivan and Affinito, Fabio and Baroni, Stefano and Baseggio, Oscar and Bellentani, Laura and Bertossa, Riccardo and Delugas, Pietro Davide and Ruffino, Fabrizio Ferrari and Orlandini, Sergio and Spiga, Filippo and Giannozzi, Paolo},
title = {{Quantum ESPRESSO}: One Further Step toward the Exascale},
journal = {Journal of Chemical Theory and Computation},
volume = {19},
number = {20},
pages = {6992-7006},
year = {2023},
doi = {10.1021/acs.jctc.3c00249}
}

@article{QE1,
doi = {10.1088/0953-8984/21/39/395502},
year = {2009},
month = {09},
publisher = {},
volume = {21},
number = {39},
pages = {395502},
author = {Paolo Giannozzi and Stefano Baroni and Nicola Bonini and Matteo Calandra and Roberto Car and Carlo Cavazzoni and Davide Ceresoli and Guido L Chiarotti and Matteo Cococcioni and Ismaila Dabo and Andrea Dal Corso and Stefano de Gironcoli and Stefano Fabris and Guido Fratesi and Ralph Gebauer and Uwe Gerstmann and Christos Gougoussis and Anton Kokalj and Michele Lazzeri and Layla Martin-Samos and Nicola Marzari and Francesco Mauri and Riccardo Mazzarello and Stefano Paolini and Alfredo Pasquarello and Lorenzo Paulatto and Carlo Sbraccia and Sandro Scandolo and Gabriele Sclauzero and Ari P Seitsonen and Alexander Smogunov and Paolo Umari and Renata M Wentzcovitch},
title = {{QUANTUM ESPRESSO}: a modular and open-source software project for quantum
simulations of materials},
journal = {Journal of Physics: Condensed Matter},
}

@article{QE2,
doi = {10.1088/1361-648X/aa8f79},
year = {2017},
month = {10},
publisher = {IOP Publishing},
volume = {29},
number = {46},
pages = {465901},
author = {P Giannozzi and O Andreussi and T Brumme and O Bunau and M Buongiorno Nardelli and M Calandra and R Car and C Cavazzoni and D Ceresoli and M Cococcioni and N Colonna and I Carnimeo and A Dal Corso and S de Gironcoli and P Delugas and R A DiStasio and A Ferretti and A Floris and G Fratesi and G Fugallo and R Gebauer and U Gerstmann and F Giustino and T Gorni and J Jia and M Kawamura and H-Y Ko and A Kokalj and E Küçükbenli and M Lazzeri and M Marsili and N Marzari and F Mauri and N L Nguyen and H-V Nguyen and A Otero-de-la-Roza and L Paulatto and S Poncé and D Rocca and R Sabatini and B Santra and M Schlipf and A P Seitsonen and A Smogunov and I Timrov and T Thonhauser and P Umari and N Vast and X Wu and S Baroni},
title = {Advanced capabilities for materials modelling with {Quantum ESPRESSO}},
journal = {Journal of Physics: Condensed Matter},
}

@article{QEGPU,
    author = {Giannozzi, Paolo and Baseggio, Oscar and Bonfà, Pietro and Brunato, Davide and Car, Roberto and Carnimeo, Ivan and Cavazzoni, Carlo and de Gironcoli, Stefano and Delugas, Pietro and Ferrari Ruffino, Fabrizio and Ferretti, Andrea and Marzari, Nicola and Timrov, Iurii and Urru, Andrea and Baroni, Stefano},
    title = "{Quantum ESPRESSO toward the exascale}",
    journal = {The Journal of Chemical Physics},
    volume = {152},
    number = {15},
    pages = {154105},
    year = {2020},
    month = {04},
    issn = {0021-9606},
    doi = {10.1063/5.0005082},
}

@misc{top500,
  howpublished =  {\url{http://www.top500.org}},
  title =         {The {TOP500} List},
  note = {Accessed: 2024-03-13}
}

@misc{nersc10,
  howpublished =  {\url{https://www.nersc.gov/systems/nersc-10/benchmarks/}},
  title =         {{NERSC}-10 Benchmark Suite},
  note = {Accessed: 2024-03-22}
}

@misc{coral2,
    howpublished = {\url{https://asc.llnl.gov/coral-2-benchmarks}},
    title = {{CORAL-2} Benchmarks},
    note = {Accessed: 2024-08-23}
}

@misc{maxcoe,
    howpublished = {\url{https://www.max-centre.eu/}},
    title = {{MaX3 CoE}},
    note = {Accessed: 2024-08-23}
}

@misc{coec,
    howpublished = {\url{https://coec-project.eu/}},
    title = {{CoEC}},
    note = {Accessed: 2024-08-23}
}

@article{Dongarra03HPL,
author = {Dongarra, Jack J. and Luszczek, Piotr and Petitet, Antoine},
title = {The {LINPACK} Benchmark: past, present and future},
journal = {Concurrency and Computation: Practice and Experience},
volume = {15},
number = {9},
pages = {803-820},
doi = {https://doi.org/10.1002/cpe.728},
year = {2003}
}

@misc{MaX,
  howpublished =  {\url{https://www.max-centre.eu/}},
  title =         {{MaX}: Materials at the e{X}ascale. An {EU Centre of Excellence} for
Supercomputing Applications},
  note = {Accessed: 2024-03-14}
}

@misc{MaXrepo,
  title        = {{MaX} benchmark repository},
  howpublished =  {\url{https://gitlab.com/max-centre/benchmarks/-/blob/master/Quantum_Espresso/CP/ZrO2/supercell_11layer/}},
}

@misc{ueabs2024,
  author = {{PRACE}},
  title = {{UEABS: Unified European Applications Benchmark Suite}},
  year = {2022},
  publisher = {Partnership for Advanced Computing in Europe},
  howpublished = {\url{https://repository.prace-ri.eu/git/UEABS/ueabs}},
  note = {Accessed: 2024-03-13}
}

@inproceedings{Malaya23Frontier,
author = {Malaya, Nicholas and Messer, Bronson and Glenski, Joseph and Georgiadou, Antigoni and Lietz, Justin and Gottiparthi, Kalyana and Day, Marc and Chen, Jackie and Rood, Jon and Esclapez, Lucas and White III, James and Jansen, Gustav R. and Curtis, Nicholas and Nichols, Stephen and Kurzak, Jakub and Chalmers, Noel and Freitag, Chip and Bauman, Paul and Fanfarillo, Alessandro and Budiardja, Reuben D. and Papatheodore, Thomas and Frontiere, Nicholas and Mcdougall, Damon and Norman, Matthew and Sreepathi, Sarat and Roth, Philip and Bykov, Dmytro and Wolfe, Noah and Mullowney, Paul and Eisenbach, Markus and Henry De Frahan, Marc T. and Joubert, Wayne},
title = {Experiences readying applications for Exascale},
year = {2023},
isbn = {9798400701092},
publisher = {Association for Computing Machinery},
address = {New York, NY, USA},
doi = {10.1145/3581784.3607065},
booktitle = {Proceedings of the International Conference for High Performance Computing, Networking, Storage and Analysis},
articleno = {53},
numpages = {13},
location = {Denver, CO, USA},
series = {SC '23}
}

@InProceedings{Juckeland15SPEC,
author="Juckeland, Guido
and Brantley, William
and Chandrasekaran, Sunita
and Chapman, Barbara
and Che, Shuai
and Colgrove, Mathew
and Feng, Huiyu
and Grund, Alexander
and Henschel, Robert
and Hwu, Wen-Mei W.
and Li, Huian
and M{\"u}ller, Matthias S.
and Nagel, Wolfgang E.
and Perminov, Maxim
and Shelepugin, Pavel
and Skadron, Kevin
and Stratton, John
and Titov, Alexey
and Wang, Ke
and van Waveren, Matthijs
and Whitney, Brian
and Wienke, Sandra
and Xu, Rengan
and Kumaran, Kalyan",
editor="Jarvis, Stephen A.
and Wright, Steven A.
and Hammond, Simon D.",
title="{SPEC ACCEL}: A Standard Application Suite for Measuring Hardware Accelerator Performance",
booktitle="High Performance Computing Systems. Performance Modeling, Benchmarking, and Simulation",
year="2015",
publisher="Springer International Publishing",
address="Cham",
pages="46--67",
isbn="978-3-319-17248-4",
doi="10.1007/978-3-319-17248-4_3"
}

@InProceedings{Boehm18SPEC,
author="Boehm, Swen
and Pophale, Swaroop
and Vergara Larrea, Ver{\'o}nica G.
and Hernandez, Oscar",
editor="Yokota, Rio
and Weiland, Mich{\`e}le
and Shalf, John
and Alam, Sadaf",
title="Evaluating Performance Portability of Accelerator Programming Models using {SPEC ACCEL} 1.2 Benchmarks",
booktitle="High Performance Computing",
year="2018",
publisher="Springer International Publishing",
address="Cham",
pages="711--723",
isbn="978-3-030-02465-9",
doi="10.1007/978-3-030-02465-9_51"
}

@Article{Sorokin20,
AUTHOR = {Sorokin, Aleksei and Malkovsky, Sergey and Tsoy, Georgiy and Zatsarinnyy, Alexander and Volovich, Konstantin},
TITLE = {Comparative Performance Evaluation of Modern Heterogeneous High-Performance Computing Systems {CPU}s},
JOURNAL = {Electronics},
VOLUME = {9},
YEAR = {2020},
NUMBER = {6},
ARTICLE-NUMBER = {1035},
ISSN = {2079-9292},
DOI = {10.3390/electronics9061035}
}

@InProceedings{Herten19ICEI,
author="Herten, Andreas
and Hater, Thorsten
and Klijn, Wouter
and Pleiter, Dirk",
editor="Weiland, Mich{\`e}le
and Juckeland, Guido
and Alam, Sadaf
and Jagode, Heike",
title="Performance Comparison for Neuroscience Application Benchmarks",
booktitle="High Performance Computing",
year="2019",
publisher="Springer International Publishing",
address="Cham",
pages="418--431",
isbn="978-3-030-34356-9",
doi="10.1007/978-3-030-34356-9_31"
}

@ARTICLE{Albers22beNNch,  
AUTHOR={Albers, Jasper and Pronold, Jari and Kurth, Anno Christopher and Vennemo, Stine Brekke and Haghighi Mood, Kaveh and Patronis, Alexander and Terhorst, Dennis and Jordan, Jakob and Kunkel, Susanne and Tetzlaff, Tom and Diesmann, Markus and Senk, Johanna},   	 
TITLE={A Modular Workflow for Performance Benchmarking of Neuronal Network Simulations}, 	
JOURNAL={Frontiers in Neuroinformatics}, 	
VOLUME={16},	
YEAR={2022},
DOI={10.3389/fninf.2022.837549}, 	
ISSN={1662-5196}
}

@INPROCEEDINGS {Farrell21MLPerf,
author = {S. Farrell and M. Emani and J. Balma and L. Drescher and A. Drozd and A. Fink and G. Fox and D. Kanter and T. Kurth and P. Mattson and D. Mu and A. Ruhela and K. Sato and K. Shirahata and T. Tabaru and A. Tsaris and J. Balewski and B. Cumming and T. Danjo and J. Domke and T. Fukai and N. Fukumoto and T. Fukushi and B. Gerofi and T. Honda and T. Imamura and A. Kasagi and K. Kawakami and S. Kudo and A. Kuroda and M. Martinasso and S. Matsuoka and H. Mendonca and K. Minami and P. Ram and T. Sawada and M. Shankar and T. t. John and A. Tabuchi and V. Vishwanath and M. Wahib and M. Yamazaki and J. Yin},
booktitle = {2021 IEEE/ACM Workshop on Machine Learning in High Performance Computing Environments (MLHPC)},
title = {{MLPerf}™ {HPC}: A Holistic Benchmark Suite for Scientific Machine Learning on {HPC} Systems},
year = {2021},
volume = {},
issn = {},
pages = {33-45},
doi = {10.1109/MLHPC54614.2021.00009},
publisher = {IEEE Computer Society},
address = {Los Alamitos, CA, USA},
month = {11}
}

@InProceedings{Xu15NAS,
author="Xu, Rengan
and Tian, Xiaonan
and Chandrasekaran, Sunita
and Yan, Yonghong
and Chapman, Barbara",
editor="Brodman, James
and Tu, Peng",
title="{NAS} Parallel Benchmarks for {GPGPUs} Using a Directive-Based Programming Model",
booktitle="Languages and Compilers for Parallel Computing",
year="2015",
publisher="Springer International Publishing",
address="Cham",
pages="67--81",
isbn="978-3-319-17473-0",
doi="10.1007/978-3-319-17473-0_5"
}

@inbook{Dongarra12HPCChallenge,
  title={{HPC} Challenge: Design, History, and Implementation Highlights},
  author={Dongarra, Jack and Luszczek, Piotr},
  editor={Vetter, Jeffrey S.},
  booktitle={Contemporary High Performance Computing: From Petascale toward Exascale},
  volume={1},
  chapter={2},
  publisher={Chapman and Hall/CRC},
  year={2013},
  isbn={9781466568358},
}

@techreport{Stratton12Parboil,
  author = {John A. Stratton and Christopher I. Rodrigues and I-Jui Sung and Nady Obeid and Li-Wen Chang and Nasser Anssari and Geng Liu and Wen-mei W. Hwu},
  title = {Parboil: A Revised Benchmark Suite for Scientific and Commercial Throughput Computing},
  year = {2012},
  institution = {University of Illinois at Urbana-Champaign, Center for Reliable and High-Performance Computing},
  number = {IMPACT-12-01},
  type = {Technical Report},
  url = {http://impact.crhc.illinois.edu/Shared/Docs/impact-12-01.parboil.pdf}
}

@inproceedings{Danalis10SHOC,
author = {Danalis, Anthony and Marin, Gabriel and McCurdy, Collin and Meredith, Jeremy S. and Roth, Philip C. and Spafford, Kyle and Tipparaju, Vinod and Vetter, Jeffrey S.},
title = {The Scalable Heterogeneous Computing ({SHOC}) benchmark suite},
year = {2010},
isbn = {9781605589350},
publisher = {Association for Computing Machinery},
address = {New York, NY, USA},
doi = {10.1145/1735688.1735702},
booktitle = {Proceedings of the 3rd Workshop on General-Purpose Computation on Graphics Processing Units},
pages = {63–74},
numpages = {12},
location = {Pittsburgh, Pennsylvania, USA},
series = {GPGPU-3}
}

@INPROCEEDINGS{Che09Rodinia,
  author={Che, Shuai and Boyer, Michael and Meng, Jiayuan and Tarjan, David and Sheaffer, Jeremy W. and Lee, Sang-Ha and Skadron, Kevin},
  booktitle={2009 IEEE International Symposium on Workload Characterization (IISWC)}, 
  title={Rodinia: A benchmark suite for heterogeneous computing}, 
  year={2009},
  volume={},
  number={},
  pages={44-54},
  doi={10.1109/IISWC.2009.5306797}}

@inproceedings{Bienia08PARSEC,
author = {Bienia, Christian and Kumar, Sanjeev and Singh, Jaswinder Pal and Li, Kai},
title = {The {PARSEC} benchmark suite: characterization and architectural implications},
year = {2008},
isbn = {9781605582825},
publisher = {Association for Computing Machinery},
address = {New York, NY, USA},
doi = {10.1145/1454115.1454128},
booktitle = {Proceedings of the 17th International Conference on Parallel Architectures and Compilation Techniques},
pages = {72–81},
numpages = {10},
location = {Toronto, Ontario, Canada},
series = {PACT '08}
}

@inproceedings{Pearce23Benchpark,
author = {Pearce, Olga and Scott, Alec and Becker, Gregory and Haque, Riyaz and Hanford, Nathan and Brink, Stephanie and Jacobsen, Doug and Poxon, Heidi and Domke, Jens and Gamblin, Todd},
title = {Towards Collaborative Continuous Benchmarking for {HPC}},
year = {2023},
isbn = {9798400707858},
publisher = {Association for Computing Machinery},
address = {New York, NY, USA},
doi = {10.1145/3624062.3624135},
booktitle = {Proceedings of the SC '23 Workshops of The International Conference on High Performance Computing, Network, Storage, and Analysis},
pages = {627–635},
numpages = {9},
location = {Denver, CO, USA},
series = {SC-W '23}
}

@Article{Fehr16RRR,
title = {Best practices for replicability, reproducibility and reusability of computer-based experiments exemplified by model reduction software},
journal = {AIMS Mathematics},
volume = {1},
number = {3},
pages = {261-281},
year = {2016},
issn = {2473-6988},
doi = {10.3934/Math.2016.3.261},
author = {Jörg Fehr and  Jan Heiland and  Christian Himpe and  Jens Saak}
}

@misc{dao2023flashattention2,
      title={{FlashAttention-2}: Faster Attention with Better Parallelism and Work Partitioning}, 
      author={Tri Dao},
      year={2023},
      eprint={2307.08691},
      archivePrefix={arXiv},
      primaryClass={cs.LG}
}

@misc{korthikanti2022reducing,
      title={Reducing Activation Recomputation in Large Transformer Models}, 
      author={Vijay Korthikanti and Jared Casper and Sangkug Lym and Lawrence McAfee and Michael Andersch and Mohammad Shoeybi and Bryan Catanzaro},
      year={2022},
      eprint={2205.05198},
      archivePrefix={arXiv},
      primaryClass={cs.LG}
}

@inproceedings{vaswani2023attention,
 author = {Vaswani, Ashish and Shazeer, Noam and Parmar, Niki and Uszkoreit, Jakob and Jones, Llion and Gomez, Aidan N and Kaiser, \L ukasz and Polosukhin, Illia},
 booktitle = {Advances in Neural Information Processing Systems},
 editor = {I. Guyon and U. Von Luxburg and S. Bengio and H. Wallach and R. Fergus and S. Vishwanathan and R. Garnett},
 pages = {},
 publisher = {Curran Associates, Inc.},
 title = {Attention is All you Need},
  url = {https://proceedings.neurips.cc/paper_files/paper/2017/file/3f5ee243547dee91fbd053c1c4a845aa-Paper.pdf},
 volume = {30},
 year = {2017}
}

@inproceedings{rajbhandari2020zero,
author = {Rajbhandari, Samyam and Rasley, Jeff and Ruwase, Olatunji and He, Yuxiong},
title = {{ZeRO}: memory optimizations toward training trillion parameter models},
year = {2020},
isbn = {9781728199986},
publisher = {IEEE Press},
booktitle = {Proceedings of the International Conference for High Performance Computing, Networking, Storage and Analysis},
articleno = {20},
numpages = {16},
location = {Atlanta, Georgia},
series = {SC '20}
}

@misc{shoeybi2020megatronlm,
      title={{Megatron-LM}: Training Multi-Billion Parameter Language Models Using Model Parallelism}, 
      author={Mohammad Shoeybi and Mostofa Patwary and Raul Puri and Patrick LeGresley and Jared Casper and Bryan Catanzaro},
      year={2020},
      eprint={1909.08053},
      archivePrefix={arXiv},
      primaryClass={cs.CL}
}

@inproceedings{narayanan2021efficient,
author = {Narayanan, Deepak and Shoeybi, Mohammad and Casper, Jared and LeGresley, Patrick and Patwary, Mostofa and Korthikanti, Vijay and Vainbrand, Dmitri and Kashinkunti, Prethvi and Bernauer, Julie and Catanzaro, Bryan and Phanishayee, Amar and Zaharia, Matei},
title = {Efficient large-scale language model training on {GPU} clusters using {Megatron-LM}},
year = {2021},
isbn = {9781450384421},
publisher = {Association for Computing Machinery},
address = {New York, NY, USA},
doi = {10.1145/3458817.3476209},
booktitle = {Proceedings of the International Conference for High Performance Computing, Networking, Storage and Analysis},
articleno = {58},
numpages = {15},
location = {St. Louis, Missouri},
series = {SC '21}
}

@INPROCEEDINGS{John2023OGX,
      author       = {John, Chelsea Maria and Ebert, Jan and Penke, Carolin and
                      Kesselheim, Stefan and Herten, Andreas},
      title        = {{O}pen{GPT}-{X} – {T}raining {L}arge {L}anguage {M}odels
                      on {HPC} {S}ystems},
      reportid     = {FZJ-2023-02173},
      year         = {2023},
      date          = {2023-05-21},
      organization  = {ISC High Performance 2023, Hamburg
                       (Germany), 21 May 2023 - 25 May 2023},
      subtyp        = {After Call},
      cin          = {JSC},
      cid          = {I:(DE-Juel1)JSC-20090406},
      pnm          = {5112 - Cross-Domain Algorithms, Tools, Methods Labs (ATMLs)
                      and Research Groups (POF4-511) / 5121 - Supercomputing $\&$
                      Big Data Facilities (POF4-512)},
      pid          = {G:(DE-HGF)POF4-5112 / G:(DE-HGF)POF4-5121},
      typ          = {PUB:(DE-HGF)24},
      doi          = {10.34732/XDVBLG-SVNDMJ},
}

@article{Edwards:2004sx,
	author        = {Edwards, Robert G. and Joo, Balint},
	editor        = {Bodwin, Geoffrey T. and Sinclair, D. K. and Eichten, E. and Holmgren, D. and Kronfeld, Andreas S. and Mackenzie, P. and Okamoto, M. and Simone, J. and El-Khadra, Aida X.},
	collaboration = {SciDAC, LHPC, UKQCD},
	title         = {The {C}hroma software system for {L}attice {QCD}},
	reportNumber  = {JLAB-THY-04-54},
	doi           = {10.1016/j.nuclphysbps.2004.11.254},
	journal       = {Nucl. Phys. B Proc. Suppl.},
	volume        = {140},
	pages         = {832},
	year          = {2005}
}

@misc{usqcd,
  title = {{USQCD}},
  lastaccessed ={March 14, 2024},
  howpublished = {\url{https://usqcd-software.github.io/}}
}

@article{Clark:2009wm,
    author = "Clark, M. A. and Babich, R. and Barros, K. and Brower, R. C. and Rebbi, C.",
    collaboration = "QUDA",
    title = "{Solving Lattice QCD systems of equations using mixed precision solvers on GPUs}",
    primaryClass = "hep-lat",
    doi = "10.1016/j.cpc.2010.05.002",
    journal = "Comput. Phys. Commun.",
    volume = "181",
    pages = "1517--1528",
    year = "2010"
}

@article{Abraham2015,
  title = {{{GROMACS}}: {{High}} Performance Molecular Simulations through Multi-Level Parallelism from Laptops to Supercomputers},
  shorttitle = {{{GROMACS}}},
  author = {Abraham, Mark James and Murtola, Teemu and Schulz, Roland and P{\'a}ll, Szil{\'a}rd and Smith, Jeremy C. and Hess, Berk and Lindahl, Erik},
  year = {2015},
  month = sep,
  journal = {SoftwareX},
  volume = {1--2},
  pages = {19--25},
  issn = {2352-7110},
  doi = {10.1016/j.softx.2015.06.001},
}

@article{Berendsen1995,
  title = {{{GROMACS}}: {{A}} Message-Passing Parallel Molecular Dynamics Implementation},
  shorttitle = {{{GROMACS}}},
  author = {Berendsen, H. J. C. and {van der Spoel}, D. and {van Drunen}, R.},
  year = {1995},
  month = sep,
  journal = {Computer Physics Communications},
  volume = {91},
  number = {1},
  pages = {43--56},
  issn = {0010-4655},
  doi = {10.1016/0010-4655(95)00042-E},
}

@incollection{pall_tackling_2014,
  title = {Tackling Exascale Software Challenges in Molecular Dynamics Simulations with {GROMACS}},
  booktitle = {Solving Software Challenges for Exascale},
  author = {P{\'a}ll, Szil{\'a}rd and Abraham, Mark James and Kutzner, Carsten and Hess, Berk and Lindahl, Erik},
  editor = {Markidis, Stefano and Laure, Erwin},
  year = {2014},
  month = apr,
  series = {Lecture Notes in Computer Science},
  number = {8759},
  pages = {3--27},
  publisher = {Springer International Publishing},
  isbn = {978-3-319-15976-8},
  doi={10.1007/978-3-319-15976-8_1}
}

@article{Pall2020,
  title = {Heterogeneous Parallelization and Acceleration of Molecular Dynamics Simulations in {{GROMACS}}},
  author = {P{\'a}ll, Szil{\'a}rd and Zhmurov, Artem and Bauer, Paul and Abraham, Mark and Lundborg, Magnus and Gray, Alan and Hess, Berk and Lindahl, Erik},
  year = {2020},
  month = oct,
  journal = {The Journal of Chemical Physics},
  volume = {153},
  number = {13},
  pages = {134110},
  publisher = {American Institute of Physics},
  issn = {0021-9606},
  doi = {10.1063/5.0018516}
}

@software{ilharco_gabriel_2021_5143773,
  author       = {Ilharco, Gabriel and
                  Wortsman, Mitchell and
                  Wightman, Ross and
                  Gordon, Cade and
                  Carlini, Nicholas and
                  Taori, Rohan and
                  Dave, Achal and
                  Shankar, Vaishaal and
                  Namkoong, Hongseok and
                  Miller, John and
                  Hajishirzi, Hannaneh and
                  Farhadi, Ali and
                  Schmidt, Ludwig},
  title        = {OpenCLIP},
  month        = jul,
  year         = 2021,
  publisher    = {Zenodo},
  version      = {0.1},
  doi          = {10.5281/zenodo.5143773}
}

@misc{radford2021learning,
      title={Learning Transferable Visual Models From Natural Language Supervision}, 
      author={Alec Radford and Jong Wook Kim and Chris Hallacy and Aditya Ramesh and Gabriel Goh and Sandhini Agarwal and Girish Sastry and Amanda Askell and Pamela Mishkin and Jack Clark and Gretchen Krueger and Ilya Sutskever},
      year={2021},
      eprint={2103.00020},
      archivePrefix={arXiv},
      primaryClass={cs.CV}
}

@article{Hohenegger2023,
author = {Hohenegger, C. and Korn, P. and Linardakis, L. and Redler, R. and Schnur, R. and Adamidis, P. and Bao, J. and Bastin, S. and Behravesh,  M. and Bergemann, M. and Biercamp, J. and Bockelmann, H. and Brokopf, R. and Brüggemann, N. and Casaroli, L. and Chegini, F. and Datseris, G. and Esch, M. and George, G. and Giorgetta, M. and Gutjahr, O. and Haak, H. and Hanke, M. and Ilyina, T. and Jahns, T. and Jungclaus, J. and Kern, M. and Klocke, D. and Kluft, L. and Kölling, T. and Kornblueh, L. and Kosukhin, S. and Kroll, C. and Lee, J. and Mauritsen, T. and Mehlmann, C. and Mieslinger, T. and Naumann and A. K. and Paccini, L. and Peinado, A. and Praturi, D. S. and Putrasahan, D. and Rast, S. and Riddick, T. and Roeber, N. and Schmidt, H. and Schulzweida, U. and Schütte, F. and Segura, H. and Shevchenko, R. and Singh, V. and Specht, M. and Stephan, C. C. and von Storch, J.-S. and Vogel, R. and Wengel, C. and Winkler, M. and Ziemen, F. and Marotzke, J. and Stevens, B.},
year = {2023},
title = {{ICON-Sapphire}: simulating the components of the Earth System and their interactions at kilometer and subkilometer scales},
journal = {Geoscientific Model Development},
volume = {16},
number = {2},
pages = {779-811},
doi = {10.5194/gmd-2022-171}
}

@article{Korn2022,
author = {Korn, P. and Brüggemann, N. and Jungclaus, J. H. and Lorenz, S. J. and Gutjahr, O. and Haak, H. and Linardakis, L. and Mehlmann, C. and Mikolajewicz, U. and Notz, D. and Putrasahan, D. A. and Singh, V. and von Storch, J.-S. and Zhu, X. and Marotzke, J.},
title = {{ICON-O}: The Ocean Component of the {ICON} Earth System Model—Global Simulation Characteristics and Local Telescoping Capability},
journal = {Journal of Advances in Modeling Earth Systems},
volume = {14},
number = {10},
doi = {10.1029/2021MS002952},
year = {2022}
}

@article{Zaengl2015,
author = {Zängl, Günther and Reinert, Daniel and Rípodas, Pilar and Baldauf, Michael},
title = {The {ICON} ({ICO}sahedral Non-hydrostatic) modelling framework of {DWD} and {MPI-M}: Description of the non-hydrostatic dynamical core},
journal = {Quarterly Journal of the Royal Meteorological Society},
volume = {141},
number = {687},
pages = {563-579},
doi = {10.1002/qj.2378},
year = {2015}
}

@article{Giorgetta2018,
author = {Giorgetta, Marco and Brokopf, R. and Crueger, T. and Esch, M. and Fiedler, Stephanie and Helmert, J. and Hohenegger, C. and Kornblueh, L. and Köhler, Martin and Manzini, Elisa and Mauritsen, Thorsten and Nam, Christine and Raddatz, Thomas and Rast, Sebastian and Reinert, D. and Sakradzija, M. and Schmidt, Hauke and Schneck, Rainer and Schnur, Reiner and Stevens, B.},
year = {2018},
month = {06},
pages = {},
title = {{ICON-A}, the atmosphere component of the {ICON} Earth System Model. Part I: Model Description},
volume = {10},
journal = {Journal of Advances in Modeling Earth Systems},
doi = {10.1029/2017MS001242}
}

@Article{Giorgetta2022,
AUTHOR = {Giorgetta, M. A. and Sawyer, W. and Lapillonne, X. and Adamidis, P. and Alexeev, D. and Cl\'ement, V. and Dietlicher, R. and Engels, J. F. and Esch, M. and Franke, H. and Frauen, C. and Hannah, W. M. and Hillman, B. R. and Kornblueh, L. and Marti, P. and Norman, M. R. and Pincus, R. and Rast, S. and Reinert, D. and Schnur, R. and Schulzweida, U. and Stevens, B.},
TITLE = {The {ICON-A} model for direct QBO simulations on GPUs (version icon-cscs:baf28a514)},
JOURNAL = {Geoscientific Model Development},
VOLUME = {15},
YEAR = {2022},
NUMBER = {18},
PAGES = {6985--7016},
DOI = {10.5194/gmd-15-6985-2022}
}

@article{Stevens2020,
  title={The Added Value of Large-eddy and Storm-resolving Models for Simulating Clouds and Precipitation},
  author={Stevens, Bjorn and Acquistapace, Claudia and Hansen, Akio and Heinze, Rieke and Klinger, Carolin and Klocke, Daniel and Rybka, Harald and Schubotz, Wiebke and Windmiller, Julia and Adamidis, Panagiotis and others},
  journal={Journal of the Meteorological Society of Japan. Ser. II},
  volume={98},
  number={2},
  pages={395-435},
  year={2020},
  doi={10.2151/jmsj.2020-021}
}

@inproceedings{PIConGPU2013,
 author = {Bussmann, M. and Burau, H. and Cowan, T. E. and Debus, A. and Huebl, A. and Juckeland, G. and Kluge, T. and Nagel, W. E. and Pausch, R. and Schmitt, F. and Schramm, U. and Schuchart, J. and Widera, R.},
 title = {Radiative Signatures of the Relativistic {Kelvin-Helmholtz} Instability},
 booktitle = {Proceedings of the International Conference on High Performance Computing, Networking, Storage and Analysis},
 series = {SC '13},
 year = {2013},
 isbn = {978-1-4503-2378-9},
 location = {Denver, Colorado},
 pages = {5:1--5:12},
 articleno = {5},
 numpages = {12},
 doi = {10.1145/2503210.2504564},
 acmid = {2504564},
 publisher = {ACM},
 address = {New York, NY, USA},
}

@ARTICLE{PIConGPU_alpaka,

  author={Burau, Heiko and Widera, Renée and Hönig, Wolfgang and Juckeland, Guido and Debus, Alexander and Kluge, Thomas and Schramm, Ulrich and Cowan, Tomas E. and Sauerbrey, Roland and Bussmann, Michael},

  journal={IEEE Transactions on Plasma Science}, 

  title={{PIConGPU}: A Fully Relativistic Particle-in-Cell Code for a {GPU} Cluster}, 

  year={2010},

  volume={38},

  number={10},

  pages={2831-2839},

  doi={10.1109/TPS.2010.2064310}}

@INPROCEEDINGS{ZenkerAsHES2016,
  author={Zenker, Erik and Worpitz, Benjamin and Widera, René and Huebl, Axel and Juckeland, Guido and Knüpfer, Andreas and Nagel, Wolfgang E. and Bussmann, Michael},
  booktitle={2016 IEEE International Parallel and Distributed Processing Symposium Workshops (IPDPSW)}, 
  title={Alpaka -- An Abstraction Library for Parallel Kernel Acceleration}, 
  year={2016},
  volume={},
  number={},
  pages={631-640},
  doi={10.1109/IPDPSW.2016.50}}

@MasterThesis{Worpitz2015,
  author = {Benjamin Worpitz},
  title  = {Investigating performance portability of a highly scalable
            particle-in-cell simulation code on various multi-core
            architectures},
  school = {{Technische Universit{\"{a}}t Dresden}},
  month  = {09},
  year   = {2015},
  type   = {Master Thesis},
  doi    = {10.5281/zenodo.49768}
}

@software{kunkel:2018:io500,
    author = {Julian Kunkel and George Markomanolis and John Bent and Jay Lofstead},
    title = {VI4IO/io-500-dev: Zenodo Citation Release},
    month = {09},
    year = {2018},
    publisher = {Zenodo},
    version = {v1.1},
    doi = {10.5281/zenodo.1422814}
}

@software{jube,
  author       = {Breuer, Thomas and
                  Wellmann, Julia and
                  Souza Mendes Guimarães, Filipe and
                  Himmels, Carina and
                  Luehrs, Sebastian},
  title        = {{JUBE}},
  month        = nov,
  year         = 2023,
  publisher    = {Zenodo},
  version      = {REL-2.6.1},
  doi          = {10.5281/zenodo.10228432}
}

@article{juwels,
  title = {{JUWELS Cluster} and {Booster}: Exascale Pathfinder with {Modular Supercomputing Architecture} at {Juelich Supercomputing Centre}},
  volume = {7},
  ISSN = {2364-091X},
  DOI = {10.17815/jlsrf-7-183},
  journal = {Journal of large-scale research facilities JLSRF},
  publisher = {Forschungszentrum Julich,  Zentralbibliothek},
  author = {Alvarez,  Damian},
  year = {2021},
  month = oct,
  pages = {A183}
}

@article{graf2021just,
  title={{JUST}: Large-Scale Multi-Tier Storage Infrastructure at the {J{\"u}lich Supercomputing Centre}},
  author={Graf, Stephan and Mextorf, Olaf},
  journal={Journal of large-scale research facilities JLSRF},
  volume={7},
  pages={A180--A180},
  year={2021},
  doi={10.17815/jlsrf-5-172}
}

@inproceedings{hoste2012easybuild,
  title={{EasyBuild}: Building Software with Ease},
  author={Hoste, Kenneth and Timmerman, Jens and Georges, Andy and De Weirdt, Stijn},
  booktitle={2012 SC Companion: High Performance Computing, Networking Storage and Analysis},
  pages={572--582},
  year={2012},
  organization={IEEE},
  doi={10.1109/SC.Companion.2012.81}
}

@misc{LinkTest,
  howpublished =  {\url{https://gitlab.jsc.fz-juelich.de/cstao-public/linktest/}},
  title =         {{LinkTest}: communication {API} benchmarking tool},
  note = {Accessed: 2024-03-22}
}

@Article{Hokkanen2021,
  author     = {Jaro Hokkanen and Stefan Kollet and Jiri Kraus and Andreas Herten and Markus Hrywniak and Dirk Pleiter},
  journal    = {Computational Geosciences},
  title      = {Leveraging {HPC} accelerator architectures with modern techniques {\textemdash} hydrologic modeling on {GPUs} with {ParFlow}},
  year       = {2021},
  month      = {05},
  doi        = {10.1007/s10596-021-10051-4},
  publisher  = {Springer Science and Business Media {LLC}},
  readstatus = {read},
}

@ARTICLE{Adamson24Jacamar,
  author={Adamson, Ryan and Bryant, Paul and Montoya, Dave and Neel, Jeff and Palmer, Erik and Powell, Ray and Prout, Ryan and Upton, Peter},
  journal={Computing in Science \& Engineering}, 
  title={Creating Continuous Integration Infrastructure for Software Development on {DOE} {HPC} Systems}, 
  year={2024},
  volume={},
  number={},
  pages={1-9},
  doi={10.1109/MCSE.2024.3362586}}

@article{occa,
title={{OCCA: A unified approach to multi-threading languages}},
author={Medina, D. S. and St-Cyr, A. and Warburton, T.},
journal={arXiv preprint arXiv:1403.0968},
year={2014}
}

@electronic{top500nov2023,
    title={{TOP500 Nov 2023}},
    year={2023},
    month={11},
    url={https://top500.org/lists/top500/2023/11/}
}

@article{dynqcd,
  title = {Leading hadronic contribution to the muon magnetic moment from lattice QCD},
  volume = {593},
  ISSN = {1476-4687},
  url = {http://dx.doi.org/10.1038/s41586-021-03418-1},
  DOI = {10.1038/s41586-021-03418-1},
  number = {7857},
  journal = {Nature},
  publisher = {Springer Science and Business Media LLC},
  author = {Borsanyi,  Sz. and Fodor,  Z. and Guenther,  J. N. and Hoelbling,  C. and Katz,  S. D. and Lellouch,  L. and Lippert,  T. and Miura,  K. and Parato,  L. and Szabo,  K. K. and Stokes,  F. and Toth,  B. C. and Torok,  Cs. and Varnhorst,  L.},
  year = {2021},
  month = apr,
  pages = {51–55}
}

@techreport{asanovic06dwarfs,
    Author= {Asanović, Krste and Bodik, Ras and Catanzaro, Bryan Christopher and Gebis, Joseph James and Husbands, Parry and Keutzer, Kurt and Patterson, David A. and Plishker, William Lester and Shalf, John and Williams, Samuel Webb and Yelick, Katherine A.},
    Title= {The Landscape of Parallel Computing Research: A View from Berkeley},
    Year= {2006},
    Month= {12},
    Url= {http://www2.eecs.berkeley.edu/Pubs/TechRpts/2006/EECS-2006-183.html},
    Number= {UCB/EECS-2006-183}
}

@ARTICLE{MatsuokaRethinkingProxy,
  author={Matsuoka, Satoshi and Domke, Jens and Wahib, Mohamed and Drozd, Aleksandr and Chien, Andrew A. and Bair, Raymond and Vetter, Jeffrey S. and Shalf, John},
  journal={Computing in Science \& Engineering}, 
  title={Preparing for the Future—Rethinking Proxy Applications}, 
  year={2022},
  volume={24},
  number={2},
  pages={85-90},
  doi={10.1109/MCSE.2022.3153105}}

@misc{zenodo_amber,
  doi = {10.5281/ZENODO.12737559},
  %url = {https://zenodo.org/doi/10.5281/zenodo.12737559},
  author = {Haghighi Mood, Kaveh and Herten, Andreas and Achilles, Sebastian},
  title = {JUPITER Benchmark Suite: Amber},
  publisher = {Zenodo},
  year = {2024},
  copyright = {MIT License}
}

@misc{zenodo_arbor,
  doi = {10.5281/ZENODO.12737757},
  %url = {https://zenodo.org/doi/10.5281/zenodo.12737757},
  author = {Hater, Thorsten and Herten, Andreas and Badwaik, Jayesh},
  title = {JUPITER Benchmark Suite: Arbor},
  publisher = {Zenodo},
  year = {2024},
  copyright = {MIT License}
}

@misc{zenodo_chroma,
  doi = {10.5281/ZENODO.12737626},
  %url = {https://zenodo.org/doi/10.5281/zenodo.12737626},
  author = {Gregory, Eric B. and Herten, Andreas and Achilles, Sebastian},
  title = {JUPITER Benchmark Suite: Chroma-LQCD},
  publisher = {Zenodo},
  year = {2024},
  copyright = {MIT License}
}

@misc{zenodo_gromacs,
  doi = {10.5281/ZENODO.12787776},
  %url = {https://zenodo.org/doi/10.5281/zenodo.12787776},
  author = {Meinke, Jan H. and Strube, Alexandre and Herten, Andreas and Achilles, Sebastian and Haghighi Mood, Kaveh},
  title = {JUPITER Benchmark Suite: GROMACS},
  publisher = {Zenodo},
  year = {2024},
  copyright = {MIT License}
}

@misc{zenodo_icon,
  doi = {10.5281/ZENODO.12787963},
  %url = {https://zenodo.org/doi/10.5281/zenodo.12787963},
  author = {Meyer, Catrin I. and Nobre Wittwer, Nils and Römmer, Manoel and Herten, Andreas and Stein, Olaf and Bishnoi, Abhiraj and Achilles, Sebastian},
  title = {JUPITER Benchmark Suite: ICON},
  publisher = {Zenodo},
  year = {2024},
  copyright = {MIT License}
}

@misc{zenodo_juqcs,
  doi = {10.5281/ZENODO.12788076},
  %url = {https://zenodo.org/doi/10.5281/zenodo.12788076},
  author = {Willsch, Dennis and De Raedt, Hans and Herten, Andreas and Achilles, Sebastian},
  title = {JUPITER Benchmark Suite: JUQCS},
  publisher = {Zenodo},
  year = {2024},
  copyright = {MIT License}
}

@misc{zenodo_nekrs,
  doi = {10.5281/ZENODO.12788254},
  %url = {https://zenodo.org/doi/10.5281/zenodo.12788254},
  author = {Witzler, Christian and Windgassen, Jonathan and Bode, Mathis and Herten, Andreas and Achilles, Sebastian},
  title = {JUPITER Benchmark Suite: nekRS},
  publisher = {Zenodo},
  year = {2024},
  copyright = {MIT License}
}

@misc{zenodo_parflow,
  doi = {10.5281/ZENODO.12788364},
  %url = {https://zenodo.org/doi/10.5281/zenodo.12788364},
  author = {Gonzalez-Nicolas, Ana and Caviedes-Voullième, Daniel and Strube, Alexandre and Achilles, Sebastian and Zhukov, Ilya and Benke, Jörg and Herten, Andreas},
  title = {JUPITER Benchmark Suite: ParFlow},
  publisher = {Zenodo},
  year = {2024},
  copyright = {MIT License}
}

@misc{zenodo_picongpu,
  doi = {10.5281/ZENODO.12788381},
  %url = {https://zenodo.org/doi/10.5281/zenodo.12788381},
  author = {Sinha, Ujjwal and Badwaik, Jayesh and Herten, Andreas and Achilles, Sebastian},
  title = {JUPITER Benchmark Suite: PIConGPU},
  publisher = {Zenodo},
  year = {2024},
  copyright = {MIT License}
}

@misc{zenodo_qe,
  doi = {10.5281/ZENODO.12788398},
  %url = {https://zenodo.org/doi/10.5281/zenodo.12788398},
  author = {Haghighi Mood, Kaveh and Herten, Andreas and Achilles, Sebastian},
  title = {JUPITER Benchmark Suite: Quantum ESPRESSO},
  publisher = {Zenodo},
  year = {2024},
  copyright = {MIT License}
}

@misc{zenodo_soma,
  doi = {10.5281/ZENODO.12788506},
  %url = {https://zenodo.org/doi/10.5281/zenodo.12788506},
  author = {Badwaik, Jayesh and Smolenko, Andreas and Herten, Andreas and Achilles, Sebastian and Breuer, Thomas},
  title = {JUPITER Benchmark Suite: SOMA},
  publisher = {Zenodo},
  year = {2024},
  copyright = {MIT License}
}

@misc{zenodo_mmoclip,
  doi = {10.5281/ZENODO.12788226},
  %url = {https://zenodo.org/doi/10.5281/zenodo.12788226},
  author = {Cherti, Mehdi and Achilles, Sebastian and Herten, Andreas and John, Chelsea and Badwaik, Jayesh},
  title = {JUPITER Benchmark Suite: MMoCLIP},
  publisher = {Zenodo},
  year = {2024},
  copyright = {MIT License}
}

@misc{zenodo_megatron,
  doi = {10.5281/ZENODO.12788115},
  %url = {https://zenodo.org/doi/10.5281/zenodo.12788115},
  author = {John, Chelsea and Kesselheim, Stefan and Penke, Carolin and Ebert, Jan and Nassyr, Stepan and Herten, Andreas and Achilles, Sebastian},
  title = {JUPITER Benchmark Suite: Megatron-LM},
  publisher = {Zenodo},
  year = {2024},
  copyright = {MIT License}
}

@misc{zenodo_resnet,
  doi = {10.5281/ZENODO.12788436},
  %url = {https://zenodo.org/doi/10.5281/zenodo.12788436},
  author = {John, Chelsea and Strube, Alexandre and Kesselheim, Stefan and Herten, Andreas and Achilles, Sebastian and Ebert, Jan},
  title = {JUPITER Benchmark Suite: ResNet},
  publisher = {Zenodo},
  year = {2024},
  copyright = {MIT License}
}

@misc{zenodo_dynqcd,
  doi = {10.5281/ZENODO.12737680},
  %url = {https://zenodo.org/doi/10.5281/zenodo.12737680},
  author = {Szabo, Kalman and Herten, Andreas and Achilles, Sebastian and Badwaik, Jayesh and Nassyr, Stepan},
  title = {JUPITER Benchmark Suite: DynQCD},
  publisher = {Zenodo},
  year = {2024},
  copyright = {MIT License}
}

@misc{zenodo_nastja,
  doi = {10.5281/ZENODO.12788527},
  %url = {https://zenodo.org/doi/10.5281/zenodo.12788527},
  author = {Behle, Eric and Herten, Andreas and Achilles, Sebastian},
  title = {JUPITER Benchmark Suite: NAStJA},
  publisher = {Zenodo},
  year = {2024},
  copyright = {MIT License}
}

@misc{zenodo_graph500,
  doi = {10.5281/ZENODO.12788553},
  %url = {https://zenodo.org/doi/10.5281/zenodo.12788553},
  author = {Alvarez, Damian and Strube, Alexandre and Herten, Andreas and Achilles, Sebastian},
  title = {JUPITER Benchmark Suite: Graph500},
  publisher = {Zenodo},
  year = {2024},
  copyright = {MIT License}
}

@misc{zenodo_hpcg,
  doi = {10.5281/ZENODO.12788585},
  %url = {https://zenodo.org/doi/10.5281/zenodo.12788585},
  author = {Mirus, Jan-Oliver and Breuer, Thomas and Achilles, Sebastian and Herten, Andreas},
  title = {JUPITER Benchmark Suite: HPCG},
  publisher = {Zenodo},
  year = {2024},
  copyright = {MIT License}
}

@misc{zenodo_hpl,
  doi = {10.5281/ZENODO.12788612},
  %url = {https://zenodo.org/doi/10.5281/zenodo.12788612},
  author = {Achilles, Sebastian and Herten, Andreas and Haghighi Mood, Kaveh},
  title = {JUPITER Benchmark Suite: HPL},
  publisher = {Zenodo},
  year = {2024},
  copyright = {MIT License}
}

@misc{zenodo_linktest,
  doi = {10.5281/ZENODO.12788705},
  %url = {https://zenodo.org/doi/10.5281/zenodo.12788705},
  author = {Müller, Yannik and Holicki, Max and Herten, Andreas and Achilles, Sebastian},
  title = {JUPITER Benchmark Suite: LinkTest},
  publisher = {Zenodo},
  year = {2024},
  copyright = {MIT License}
}

@misc{zenodo_osu,
  doi = {10.5281/ZENODO.12788751},
  %url = {https://zenodo.org/doi/10.5281/zenodo.12788751},
  author = {Breuer, Thomas and Herten, Andreas and Achilles, Sebastian},
  title = {JUPITER Benchmark Suite: OSU Micro-Benchmarks},
  publisher = {Zenodo},
  year = {2024},
  copyright = {MIT License}
}

@misc{zenodo_stream,
  doi = {10.5281/ZENODO.12788781},
  %url = {https://zenodo.org/doi/10.5281/zenodo.12788781},
  author = {Achilles, Sebastian and Breuer, Thomas and Thust, Kay and Müller, Yannik and Herten, Andreas and Strube, Alexandre and Krause, Dorian and El Sayed, Salem},
  title = {JUPITER Benchmark Suite: STREAM},
  publisher = {Zenodo},
  year = {2024},
  copyright = {MIT License}
}

@misc{zenodo_streamgpu,
  doi = {10.5281/ZENODO.12788801},
  %url = {https://zenodo.org/doi/10.5281/zenodo.12788801},
  author = {Achilles, Sebastian and Herten, Andreas and Haghighi Mood, Kaveh},
  title = {JUPITER Benchmark Suite: STREAM},
  publisher = {Zenodo},
  year = {2024},
  copyright = {MIT License}
}
\end{document}